\documentclass[fleqn,usenatbib]{mnras}

\usepackage{newtxtext,newtxmath}
\usepackage{caption}

\usepackage[T1]{fontenc}

\DeclareRobustCommand{\VAN}[3]{#2}
\let\VANthebibliography\thebibliography
\def\thebibliography{\DeclareRobustCommand{\VAN}[3]{##3}\VANthebibliography}

\usepackage{graphicx}	
\usepackage{amsmath}	

\usepackage{hyperref}

\title{Pre-main Sequence B Stars Surrounding the Orion Nebula}

\author[P.H.F.B. Braz et al.]{
P.H.F.B. Braz,$^{1}$\thanks{E-mail: pedrobbraz@ufmg.br}
A. Roman-Lopes,$^{2}$
W.J.B. Corradi,$^{3,1}$
E. B. Amôres,$^{4,1}$
M. S. Angelo,$^{5}$
F. S. S. Maia,$^{6}$
\newauthor
J. F. C. Santos Jr.,$^{1}$
A. E. Piatti$^{7,8}$
and W. Reis$^{9}$
\\
$^{1}$Departamento de F\'isica, ICEx, Universidade Federal de Minas Gerais, Av. Ant\^onio Carlos 6627, 31270-901 Belo Horizonte, MG, Brazil\\
$^{2}$Department of Astronomy, Universidad de La Serena, Av. Raul Bitran \#1302, La Serena, Chile\\
$^{3}$Laborat\'orio Nacional de Astrof\'isica, R. Estados Unidos, 154, 37504-364, Itajub\'a, MG, Brazil\\
$^{4}$Departamento de F\'isica, Universidade Estadual de Feira de Santana (UEFS), Av. Transnordestina, S/N, CEP 44036-900 Feira de Santana, BA, Brazil\\
$^{5}$Centro Federal de Educa\c{c}\~ao Tecnol\'ogica de Minas Gerais, Av. Amazonas, 7675, 30510-000 Belo Horizonte, MG, Brazil\\
$^{6}$Instituto de F\'isica, Universidade Federal do Rio de Janeiro, 21941-972 Rio de Janeiro, Brazil\\
$^{7}$Instituto Interdisciplinario de Ciencias B\'asicas (ICB), CONICET-UNCuyo, Padre J. Contreras 1300, M5502JMA, Mendoza, Argentina\\
$^{8}$Consejo Nacional de Investigaciones Cient\'{\i}ficas y T\'ecnicas, Godoy Cruz 2290, C1425FQB,  Buenos Aires, Argentina\\
$^{9}$IBMEC, R. Rio Grande do Norte 300, 30130-130, Belo Horizonte, MG, Brazil
}

\date{Accepted XXX. Received YYY; in original form ZZZ}

\pubyear{2025}

\begin{document}
\label{firstpage}
\pagerange{\pageref{firstpage}--\pageref{lastpage}}
\maketitle

\begin{abstract}
Pre-main sequence (PMS) early-type stars are rare due to their rapid evolution. Confirming PMS B stars is accordingly valuable for constraining higher mass star formation scenarios. Although the Orion Nebula (ON) region offers an ideal laboratory for such studies, its population of B stars remains poorly characterized. We aim to determine spectral types, masses, and ages of B and early A stars surrounding the ON, to identify robust PMS candidates. We combine optical and near-infrared spectroscopy to estimate spectral types using Brackett, He, Si, and Mg lines. Photometry from \textit{Gaia}, 2MASS, and WISE is used to construct colour-magnitude diagrams to fit isochrones via $\chi^2$ minimisation, yielding stellar ages and masses. Dynamical masses from eclipsing binaries are employed as independent constraints on the mass range allowed for each spectral subtype. We derive stellar ages and masses for 53 stars and spectral classifications for 48 of them, identifying 30 PMS candidates, including some that were previously assigned to luminosity classes I–III. Several stars show spectroscopic variability, potentially linked to circumstellar material or binarity. Our combined spectroscopic and photometric approach identifies robust PMS B-star candidates and provides a validated framework for distinguishing them from evolved counterparts -- in this instance, refining the census of early-types young stars in Orion. The age distribution of the PMS candidates offers preliminary clues about the star formation history of the ON region.
\end{abstract}

\begin{keywords}
Star: Early-type -- Star: Formation -- Star: Pre-Main Sequence
\end{keywords}



\section{Introduction}

    Early-type stars (spectral type B or earlier) play a fundamental role in the evolution of galaxies by injecting energy, momentum, and heavy elements into the interstellar medium through strong stellar winds, ionising radiation, and supernova explosions \citep{zinnecker07,krumholz14}. Their rapid evolutionary timescales imply that the pre-main sequence (PMS) phase is extremely short-lived, making the direct detection of very young B stars a challenging but important task \citep{Ochsendorf11,ramirezTannus17,motte18}. 

    Several well-known features, such as emission lines -- in particular H$\alpha$ line -- infrared (IR) excess, and photometric variability are often used to identify intermediate-mass PMS stars, as these characteristics are indicative of relic circumstellar dust disc \citep{The94,FinkenzellerMundt84,vieira03,Vioque22}. Stars in the PMS phase in the mass range of about 2 to 10\,M$_\odot$, presenting a circumstellar disc, are the so-called Herbig Ae/Be stars \citep{herbig60,Brittain23}. They commonly display spectroscopic variability as evidence of interaction between the star and its accretion disc, which can be caused by infalling (accretion mechanisms) or outflowing matter (ejection mechanism) \citep[][Braz et al. in prep.]{Guimaraes06,Pogodin21,Derkink24}.  
    
    However, not all intermediate-mass PMS stars show evidence of a circumstellar disc, as they can be rapidly evaporated due to the radiation of the central star, leading to a very short lifetime ($\sim10^5$\,yr) \citep{Natta00,AlonsoAlbi09}. In this sense, to investigate the properties of PMS B stars, it is essential to constrain the models of more massive star formation, since these objects provide a connection to establish a proper comprehension of the high-mass and the (relatively well-known) low-mass star formation. Moreover, since these objects are expected to remain close to their natal environment and retain key spectroscopic and photometric imprints of their early evolution, they can be used to trace the star formation history of the region where they were born. 
    
    Among the various star-forming regions, the Orion complex stands out as a benchmark laboratory due to its proximity ($\sim$400 pc), multi-scale structure, and ongoing star formation spanning from low- to high-mass regimes \citep{bally08,megeath16}. While extensive efforts have addressed the large-scale star formation history of the Orion complex \citep[e.g.][]{bally08,bouy14,kounkel18,grossschedl21}, and the characterization of its later-type young stellar population \citep[e.g.][]{Grossschedl19,Zari19,Wright24,SanchezSanjuan24}, the properties of early-type stars in specific substructures — such as the immediate surroundings of the Orion Nebula (ON) — remain comparatively underexplored.
    
    Luminosity classes I–III have been assigned to several B-type stars near the ON \citep{houk99}. However, PMS stars can mimic these spectral characteristics, leading to potential misclassification \citep{graycorbally09,sota11,ramirez20}. Disentangling PMS B stars from evolved counterparts requires precise spectral type determination — ideally combining optical and infrared (IR) diagnostics — and independent constraints on their physical parameters.
    
    The evolutionary stage of a star can be further constrained by its position in a colour–magnitude diagram (CMD) compared to theoretical isochrones \citep[e.g.][]{bre12}. Spectral type estimates place strong limits on the stellar mass, reducing degeneracies in age determination. This synergy enables a self-consistent link between mass and age during the isochrone fitting procedure, increasing the reliability of PMS identification, especially when combined with extinction corrections and metallicity constraints.
    
    In this work, we present a combined optical and IR spectroscopic study of B and early A-type stars in the vicinity of the ON. We determine their spectral types, and subsequently their masses and ages, through $\chi^2$ minimisation using PARSEC isochrones \citep{bre12}. To validate our age–mass estimates, we employed independent constraints from dynamical mass determinations for eclipsing binary systems \citep{eker18}. 
    
    This approach allowed us to reclassify objects previously assigned to evolved luminosity classes (class I-III), and to identify a robust sample of young stellar candidates in the PMS phase. We observed multiple stellar populations in different evolutionary stages, which provide insights about the star formation scenario for the early-type stellar population in the surroundings of the ON region. Finally, we discuss a subset of objects displaying peculiar features, such as spectroscopic variability, highlighting their potential as laboratories for studying early stellar evolution. 

    \begin{figure*}
        \centering
        \includegraphics[scale=0.45]{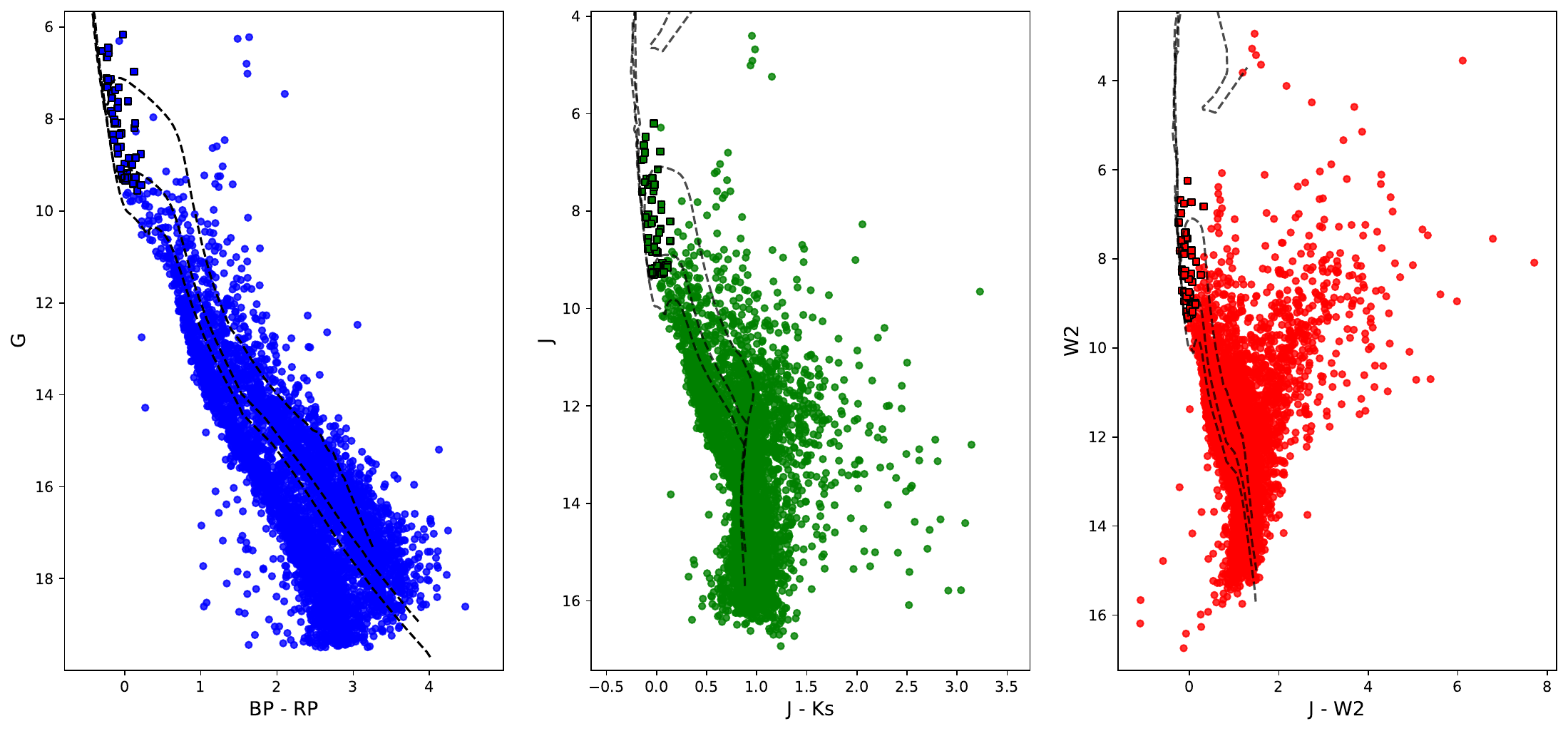}
        \includegraphics[scale=0.5]{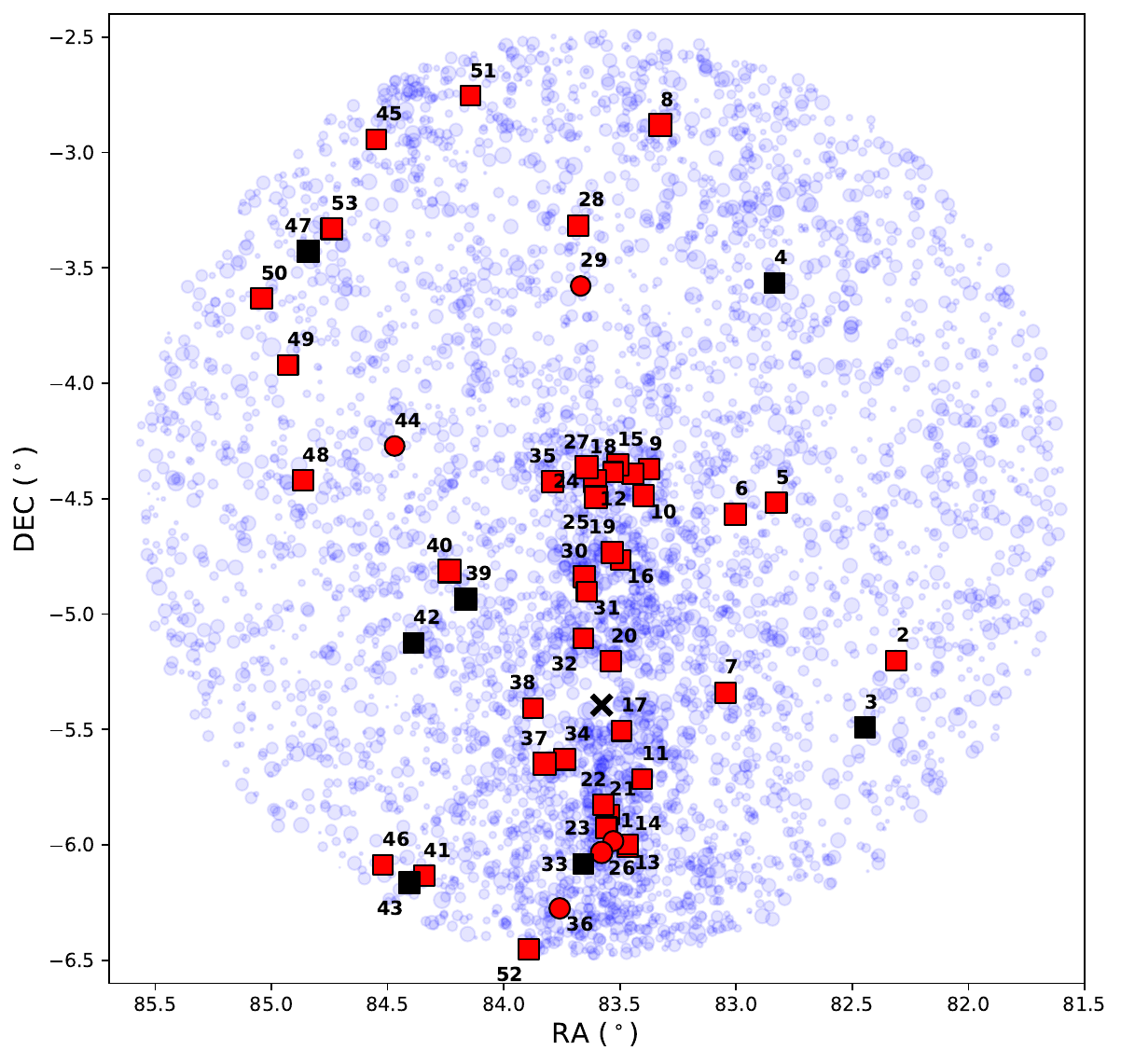}
        \caption{CMDs (top panels) and skymap (bottom panel) of the sample selection from the \textit{Gaia}, 2MASS, and WISE crossmatch, considering sources with $2 \leq \varpi \leq 3$\,mas located within 2$^\circ$ of the ON. The top panels show the \textit{Gaia}, 2MASS and WISE CMDs, respectively. In each CMD, the early-type candidates selected in this work are outlined in black, while the dashed lines indicate the 1, 5, and 10 Myr isochrones for visual guidance. In the skymap, the sizes of all symbols are proportional to the G magnitude, with the blue circles representing the photometric crossmatched sample, whereas the squares and circles denote the early-type candidates with and without available spectra, respectively. Runaway candidates are shown in black. The number near each star corresponds to its index in Table\,\ref{tab:spec_class}. The black X marks the centre of the ON cluster as determined by Kuhn et al. (2019).}
        \label{fig:skymap_crossmatch}
    \end{figure*}

    The paper is organized as follows. In Sect.\,\ref{sec:star_selection}, we present the data and sample selection. The methodology used to estimate the spectral types is detailed in Sect.\,\ref{sec:spectral_classification}, while Sect.\,\ref{sec:age_mass_estimation} describes the age and mass estimation and the comparison with the masses from eclipsing binary systems. The PMS candidates and their age distribution are presented in Sect.\,\ref{sec:pms_candidates}. Finally, the peculiar features of the sample are briefly discussed in Sect.\,\ref{Sect:special_cases}, and the conclusions are presented in Sect.\,\ref{sec:conclusions}.

\section{Selection of B and A stars surrounding the ON}
    \label{sec:star_selection}
    
    To construct the CMDs and define our target sample, we used astrometric and photometric data from the \textit{Gaia} DR3 \citep{gaiaDR3}, 2MASS \citep{2mass}, and WISE \citep{wise} surveys. We cross-matched the three catalogues by comparing the stars position and selecting the best match within 1". The search was restricted to a circular area of $2^\circ$ radius centred on $\textrm{RA}=83.81^\circ$ and $\textrm{DEC}=-4.50^\circ$ and to stars with parallaxes in the range $2 \leq \varpi \leq 3$\,mas, corresponding to distances of $330 \lesssim d \lesssim 500$\,pc, which was applied to include the majority of stars surrounding the ON while minimising background and foreground contamination. From the initial crossmatch between \textit{Gaia}, 2MASS, and WISE catalogues, a total of 49547 sources were obtained, while after applying the parallax filter, the sample was reduced to 4860 stars. 
    
    To focus on the brightest early-type stars (A0 or earlier), we imposed magnitude limits of $G \leq 9.6$, $J \leq 9.4$, and $W2 \leq 9.4$, along with the following colour constraints:
    \begin{eqnarray*}
       -0.3 \leq G_\mathrm{BP} - G_\mathrm{RP} \leq 0.25 \\
       -0.2 \leq J - K_\mathrm{s} \leq 0.15 \\
       -0.3 \leq J - W2 \leq 0.35 \\
    \end{eqnarray*}
    These limits were adopted by constructing CMDs for the three catalogues with their respective 1, 5, and 10\,Myrs solar-metallicity isochrones corrected only for a general distance modulus of 8.00\,mag ($\sim$400\,pc). These isochrones were used as guidelines to identify an initial sample of early-type candidates. The CMDs with these isochrones are shown in the top panels of Fig.\,\ref{fig:skymap_crossmatch}, whereas the selected early-type candidates are outlined in black. The \textit{Gaia} CMD is particularly useful for this first selection, as the early-type stars are concentrated at its bright end. 
    
    By selecting stars near the isochrones, we established an initial, broad set of colour and magnitude thresholds. We then searched for available optical and IR spectra in the APOGEE \citep{apogee} and LAMOST \citep{LAMOST} surveys or, when no spectra were available, for spectral classification in \cite{Skiff09} database. This procedure was used to verify whether stars located near the adopted limits corresponded to an early or late-type star. The colour and magnitude limits were subsequently refined until no stars of spectral type F or later remained in the sample. 
    
    Applying the magnitude filter removed late-type main-sequence (MS) stars from the sample, reducing the number of sources from 4860 to 79 stars. By applying the abovementioned colour filters, which were used to exclude evolved late-type stars, we further reduced the sample to 53 objects. Although this selection procedure proved effective at isolating early-type stars, it does not recover the entire population of B stars surrounding the ON. 
    
    To investigate the completeness of our selection, we compared our sample with the census of OB stars presented by \cite{Quintana25}. We selected from their catalogue all stars with $2 \leq \varpi \leq 3$\,mas located within 2$^\circ$ of the centre adopted in this work. This comparison yielded 30 stars in common between the two samples, while 16 early-type objects from their catalogue are absent from our sample. Most of these ``missing'' stars are deeply embedded within the ON and exhibit high extinction values, as is the case for several members of the Trapezium Cluster. Owing to the crossmatch between the three catalogues and the adopted selection criteria -- particularly the colour filters -- these highly reddened stars are not included in our sample. 
    
    On the other hand, the catalogue of \cite{Quintana25} does not fully overlap our sample, as 23 stars are present only in our catalogue. The differences between the two samples reflect the distinct selection functions and scientific goals of each work, and the two catalogues provide complementary views of the early-type stellar population surrounding the ON. The skymap of the selected early-type candidates is shown in the bottom panel of Fig.\,\ref{fig:skymap_crossmatch}, where the black X marks the ON cluster centre \citep{Kuhn19}. The early-type candidates are represented by the red and black symbols, while the blue circles correspond to all sources from the \textit{Gaia}, 2MASS and WISE crossmatch satisfying $2 \leq \varpi \leq 3$\,mas. 
    
    Using \textit{Gaia} and 2MASS IDs, we cross-matched the resulting photometric sample with the APOGEE-2 survey to obtain near-infrared spectra. For optical spectroscopy, we obtained data with the 2.15\,m Jorge Sahade Telescope at CASLEO (Argentina), coupled with Échelle REOSC (R $\sim$ 12000) and reduced using standard {\textsc IRAF} routines, and from the 8.1\,m Gemini telescope equipped with GHOST in the standard mode (R $\sim$ 50000), reduced with the DRAGONS pipeline \citep{dragons}.

    \begin{figure}
        \centering
        \includegraphics[scale=0.45]{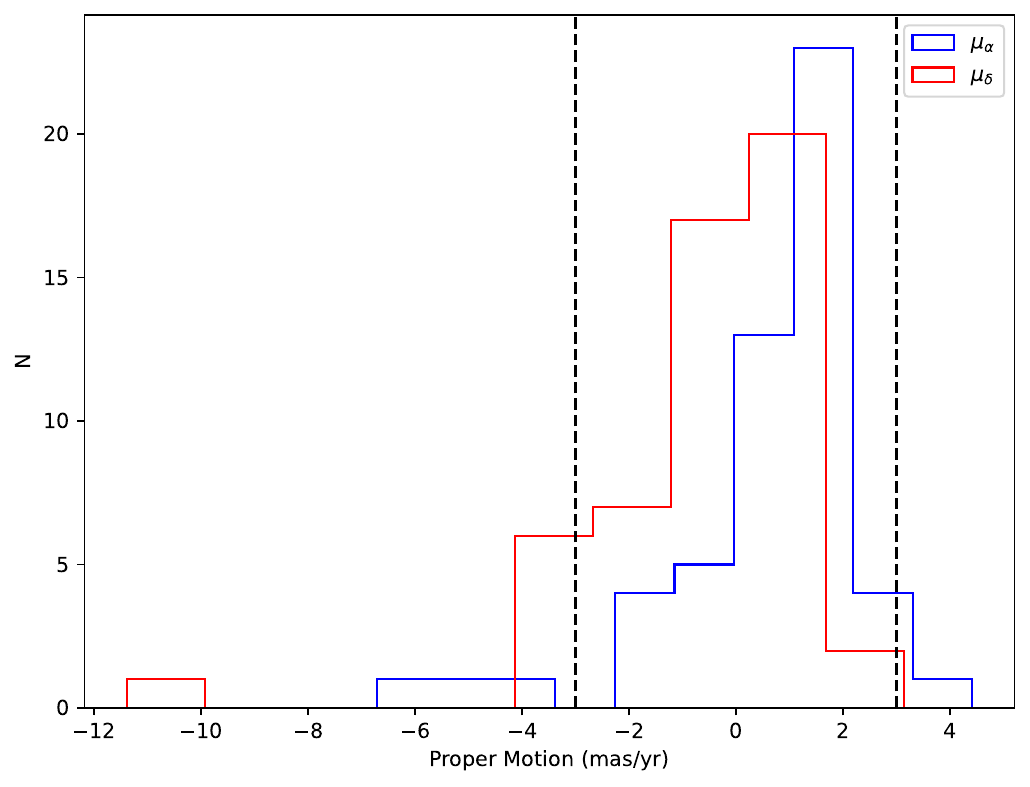}
        \caption{Histogram of $\mu_\alpha$ (blue) and $\mu_\delta$ (red) for the selected sample of early-type candidates surrounding the ON region. The dashed lines represent the adopted kinematics boundaries at $\pm$3\,mas/yr for the ON population.}
        \label{fig:hist_PM_sample_review}
        \end{figure}

    To complement the optical dataset, we retrieved spectra from the LAMOST DR10 survey and from the ESO archive. This process yielded 37 stars with IR spectra and 27 with optical spectra, giving a total of 48 stars with at least one spectroscopic observation. Additionally, five stars without available IR or optical spectra, represented by the red circles in Fig.\,\ref{fig:skymap_crossmatch}, were kept in the sample for subsequent photometric analysis. For spectral classification, we obtained spectra from the APOGEE (IR) and LAMOST (optical) surveys from selected stars classified as B2V to A0V in \citet{ramirez20} and used them as template stars.

    \subsection{Astrometric Information for the Early-type Candidates}

        \begin{table}
            \centering
            \caption{Table of runaway candidates. The ID column shows the HD number and the index in parentheses used in this work to identify the stars. The other columns show their proper motions in $\alpha$ and $\delta$.}
            \begin{tabular}{crr}\hline
                ID & $\mu_\alpha$ (mas/yr) & $\mu_\delta$ (mas/yr) \\\hline
                HD  36324 (\#3)  & $0.90 \pm 0.02$ & $-3.42 \pm 0.02$ \\
                HD  36527 (\#4)  & $-2.06 \pm 0.02$ & $-3.86 \pm 0.01$ \\
                HD  37078 (\#33) & $-3.70 \pm 0.02$ & $-11.39 \pm 0.02$ \\
                HD  37334 (\#39) & $-0.25 \pm 0.05$ & $3.14 \pm 0.04$ \\
                HD  37455 (\#42) & $-5.27 \pm 0.02$ & $-2.98 \pm 0.01$ \\
                HD  37470 (\#43) & $4.42 \pm 0.04$ & $-2.32 \pm 0.04$ \\
                HD  37687 (\#47) & $-6.72 \pm 0.08$ & $-1.13 \pm 0.08$ \\\hline
            \end{tabular}
            \label{table_PM_outliers}
        \end{table}

        Many studies have investigated the kinematic properties of the different stellar groups within the Orion Complex region \citep{kounkel18,Zari19,grossschedl21,SanchezSanjuan24}. Since our sample was crossmatched with \textit{Gaia} to use the photometric and parallax data, it is straightforward to incorporate the proper motion information. Although these previous works employed different clustering algorithms and parameter choices to separate stellar populations -- leading to differences in the proper motion limits for the ON population -- their results provide useful guidance for defining a kinematic selection criteria.

        Based on these studies, we have adopted the limit $-2.5 \leq \mu_\alpha \leq 3$ and $-3 \leq \mu_\delta \leq 3$ for stars surrounding the ON. It can be seen in Fig.\,\ref{fig:hist_PM_sample_review}, which shows the proper motion histograms of our selected sample, with reference lines at $\pm$ 3\,mas/yr indicating the adopted kinematic boundaries for stars consistent with the ON population, that nearly all our sources fall within this limit.

        The IDs and proper motions of the seven stars with distinct kinematics, hereafter referred to as runaway candidates, are listed in Table.\,\ref{table_PM_outliers}. It is worth noting that HD\,36324 (\#3), HD\,36527 (\#4), and HD\,37334 (\#39) exceed the adopted limits only marginally, by at most 0.9\,mas/yr. Moreover, except for HD\,37078 (\#33), these runaway candidates are located farther from the ON, as can be seen in Fig.\,\ref{fig:skymap_crossmatch} by the black squares, which represent the runaway candidates. Furthermore, among the seven runaway candidates, only the proper motion of \#4 points toward the ON. The remaining stars either show proper motions pointing away from the ON, or it is inconclusive to determine whether these stars were formed within the ON region. Despite their distinct proper motions, we retain these stars as kinematically peculiar sources potentially related to the broader Orion region, as their spectral classification, mass or age estimates can still be valuable information.  

    \begin{table*}
    \centering
    \renewcommand{\arraystretch}{1.3}
        \caption{Spectral type, mass and age estimation of B and A-type stars sample surrounding the Orion Nebula. The first column presents the star's identifier, the spectral type from the literature (Ref) with the corresponding reference, and those estimated in this work through infrared (IR) and optical (Opt) spectra. The spectral types marked with a ``*'' were estimated only by comparing with the template stars, while those presenting a ``+'' are stars of spectral type later than A0. The final columns show the best age (in Myrs) and mass (in M$_\odot$) estimated per survey with their respective errors.}
        \resizebox{\textwidth}{!}{
\begin{tabular}{ccccccccccc}\hline        
Index & ID & \multicolumn{3}{c}{SpType} & \multicolumn{2}{c}{\textit{Gaia}} & \multicolumn{2}{c}{2MASS} & \multicolumn{2}{c}{WISE}\\
        &           & Ref          & IR     & Opt    & Age & Mass & Age  & Mass & Age & Mass \\ \hline
1  & Brun 508  & B9V $^8$         & -  & -  &3.0$^{+8.0}_{-1.0}$&3.3 $\pm$ 0.5 &11.0$^{+0.0}_{-9.0}$ &3.3 $\pm$ 0.7 &4.0$^{+4.0}_{-2.0}$&3.5 $\pm$ 0.2 \\
2  & HD 36234  & B9III $^1$       & B8V  & -  &2.0$^{+6.0}_{-0.0}$&3.4 $\pm$ 0.5 &2.0$^{+9.0}_{-0.0}$ &3.4 $\pm$ 0.7 &2.0$^{+9.0}_{-0.0}$&3.3 $\pm$ 0.7 \\
3  & HD 36324  & A0/1IV $^1$      & B9V  & B9V  &4.0$^{+0.0}_{-1.0}$&2.2 $\pm$ 0.5 &4.0$^{+0.0}_{-1.0}$ &2.2 $\pm$ 1.1 &4.0$^{+0.0}_{-1.0}$&2.2 $\pm$ 0.4 \\

4 & HD  36527 & A0IV$^1$              & - & A0V &5.0$^{+1.0}_{-1.0}$&2.1 $\pm$ 0.2 &6.0$^{+0.0}_{-1.0}$ &1.8 $\pm$ 0.2 &6.0$^{+1.0}_{-1.0}$&1.9 $\pm$ 0.2 \\
5 & HD  36540 & B8III$^1$             & - & B7V &2.0$^{+0.0}_{-1.0}$&2.9 $\pm$ 1.2 &2.0$^{+0.0}_{-1.2}$ &2.8 $\pm$ 1.7 &2.0$^{+0.0}_{-1.0}$&2.9 $\pm$ 1.1 \\            

6  & HD 36629  & B2V $^{3,4}$     & B2V  & B2V  &1.0$^{+0.0}_{-0.0}$&4.0 $\pm$ 0.1 &1.0$^{+0.0}_{-0.1}$ &3.9 $\pm$ 0.2 &1.0$^{+0.0}_{-0.3}$&3.9 $\pm$ 0.6 \\            
7  & HD 36655  & B9V $^1$         & B9V*  & -  &2.0$^{+9.0}_{-0.0}$&3.3 $\pm$ 0.8 &3.0$^{+1.0}_{-1.0}$ &2.5 $\pm$ 0.6 &4.0$^{+0.0}_{-1.0}$&2.2 $\pm$ 0.4 \\
8  & HD 36827  & B3V$^2$          & B4V*  & -  &0.6$^{+9.4}_{-0.4}$&8.1 $\pm$ 2.5 &4.0$^{+4.0}_{-3.7}$ &8.1 $\pm$ 2.4 &4.0$^{+3.0}_{-3.8}$&9.1 $\pm$ 1.0 \\
9  & HD 36842  & B6V $^{1,4}$     & B4V  & -  &2.0$^{+9.0}_{-1.1}$&4.3 $\pm$ 0.5 &8.0$^{+1.0}_{-7.1}$ &4.3 $\pm$ 0.5 &11.0$^{+0.0}_{-10.2}$&4.5 $\pm$ 0.3 \\
10 & HD 36865A & B8/9V $^{1,4}$   & B6V  & B7V  &0.9$^{+10.1}_{-0.2}$&4.2 $\pm$ 1.3 &0.9$^{+0.1}_{-0.0}$ &4.1 $\pm$ 0.3 &0.9$^{+0.1}_{-0.3}$&4.1 $\pm$ 0.7 \\
11 & HD 36866  & A3Vas $^9$       & A0V+ & -  &5.0$^{+0.0}_{-1.0}$&2.0 $\pm$ 0.3 &5.0$^{+1.0}_{-1.0}$ &2.0 $\pm$ 0.3 &5.0$^{+1.0}_{-1.0}$&2.0 $\pm$ 0.3 \\
12 & HD 36883A & B5/7III $^1$     & B6V  & B6V  &0.9$^{+8.1}_{-0.2}$&4.3 $\pm$ 1.3 &0.9$^{+0.1}_{-0.3}$ &4.0 $\pm$ 0.8 &0.7$^{+0.2}_{-0.3}$&4.5 $\pm$ 1.4 \\
13 & HD 36918  & B5/8V $^{7,4}$   & B7V  & B7V  &3.0$^{+7.0}_{-1.0}$&3.9 $\pm$ 0.4 &2.0$^{+7.0}_{-0.0}$ &3.9 $\pm$ 0.3 &2.0$^{+1.0}_{-0.0}$&4.1 $\pm$ 0.2 \\

14 & HD  36919 & B9V$^8$              & - & B9V &4.0$^{+3.0}_{-0.0}$&2.6 $\pm$ 0.3 &4.0$^{+1.0}_{-0.0}$ &2.6 $\pm$ 0.5 &4.0$^{+2.0}_{-0.0}$&2.5 $\pm$ 0.3 \\

15 & HD 36936  & B5V $^6$         & B4V  & B6V  &0.7$^{+6.3}_{-0.2}$&5.5 $\pm$ 1.2 &0.9$^{+0.1}_{-0.2}$ &4.2 $\pm$ 0.5 &0.8$^{+9.2}_{-0.1}$&4.4 $\pm$ 1.7 \\
16 & HD 36938  & B8V $^{1,4}$     & -    & B8V  &3.0$^{+1.0}_{-0.0}$&2.6 $\pm$ 0.4 &4.0$^{+0.0}_{-1.0}$ &2.1 $\pm$ 0.4 &5.0$^{+0.0}_{-1.0}$&1.9 $\pm$ 0.2 \\
17 & HD 36939  & B8V $^4$         & B7V  &   -  &3.0$^{+8.0}_{-0.0}$&2.7 $\pm$ 0.5 &4.0$^{+1.0}_{-1.0}$ &2.2 $\pm$ 0.4 &3.0$^{+2.0}_{-1.0}$&2.6 $\pm$ 0.9 \\
18 & HD 36957  & A1 $^4$          & A0V+ & A0V  &3.0$^{+8.0}_{-1.0}$&2.6 $\pm$ 0.7 &3.0$^{+8.0}_{-1.0}$ &2.6 $\pm$ 1.1 &3.0$^{+1.0}_{-1.0}$&2.6 $\pm$ 0.8 \\
19 & HD 36958  & B3/5V $^{1,4}$   & B3V  & B4V  &0.8$^{+0.2}_{-0.4}$&4.3 $\pm$ 1.7 &0.7$^{+0.3}_{-0.2}$ &4.5 $\pm$ 0.8 &0.6$^{+0.3}_{-0.2}$&4.8 $\pm$ 0.9 \\
20 & HD 36981  & B5V $^4$         & B5V  &   -  &0.9$^{+8.1}_{-0.2}$&4.9 $\pm$ 0.6 &9.0$^{+1.0}_{-8.3}$ &4.7 $\pm$ 0.6 &1.0$^{+10.0}_{-0.4}$&5.3 $\pm$ 0.7 \\
21 & HD 36983  & B8V $^{4,10}$    & B7V  &   -  &4.0$^{+1.0}_{-1.0}$&2.7 $\pm$ 0.3 &4.0$^{+7.0}_{-1.0}$ &2.4 $\pm$ 0.7 &11.0$^{+0.0}_{-8.0}$&2.9 $\pm$ 0.6 \\
22 & HD 36999  & B8 $^4$          & B6V  & B8V  &4.0$^{+6.0}_{-2.0}$&3.7 $\pm$ 0.4 &2.0$^{+9.0}_{-0.0}$ &3.7 $\pm$ 0.4 &2.0$^{+5.0}_{-0.0}$&3.9 $\pm$ 0.8 \\
23 & HD 37000  & B4 $^4$          & B4V* & B5V  &2.0$^{+4.0}_{-1.4}$&5.8 $\pm$ 1.3 &1.0$^{+0.0}_{-0.3}$ &3.9 $\pm$ 0.8 &0.6$^{+0.4}_{-0.2}$&4.9 $\pm$ 1.2 \\
24 & HD 37016A & B8III $^1$       & B3V  & B4V  &8.0$^{+3.0}_{-7.8}$&8.1 $\pm$ 2.5 &0.2$^{+8.8}_{-0.0}$ &7.4 $\pm$ 2.9 &3.0$^{+1.0}_{-2.9}$&11.1 $\pm$ 3.8 \\

25 & HD  37017 & B2/3V$^1$            & - & B3V &3.0$^{+8.0}_{-2.8}$&8.1 $\pm$ 2.6 &1.0$^{+10.0}_{-0.8}$ &9.1 $\pm$ 3.6 &0.8$^{+10.2}_{-0.5}$&10.1 $\pm$ 3.9 \\

26 & HD 37025  & B3 $^{1,5}$      & -    & -  &11.0$^{+0.0}_{-10.6}$&6.1 $\pm$ 1.5 &9.0$^{+0.0}_{-8.7}$ &6.5 $\pm$ 2.2 &0.7$^{+0.3}_{-0.2}$&4.5 $\pm$ 0.8 \\

27 & HD  37040 & B2V$^1$          & - & B3V &8.0$^{+0.0}_{-7.8}$&8.1 $\pm$ 2.5 &10.0$^{+1.0}_{-9.7}$ &8.6 $\pm$ 2.4 &0.3$^{+3.7}_{-0.1}$&6.1 $\pm$ 3.6 \\

28 & HD 37056  & B8/9V $^1$       & B8V* & -  &3.0$^{+5.0}_{-1.0}$&3.9 $\pm$ 0.8 &3.0$^{+0.0}_{-1.0}$ &2.5 $\pm$ 0.5 &2.0$^{+1.0}_{-0.0}$&3.1 $\pm$ 0.6 \\

29 & HD 37057  & B9V$^1$              &  -   & -  &4.0$^{+1.0}_{-1.0}$&2.5 $\pm$ 0.4 &5.0$^{+1.0}_{-0.0}$ &2.0 $\pm$ 0.2 &6.0$^{+1.0}_{-1.0}$&1.9 $\pm$ 0.2 \\

30 & HD  37058 & B4V$^{4}$           & - & B2V &0.5$^{+7.5}_{-0.0}$&6.1 $\pm$ 0.8 &2.0$^{+7.0}_{-1.6}$ &6.1 $\pm$ 0.6 &4.0$^{+6.0}_{-3.6}$&6.5 $\pm$ 0.8 \\

31 & HD 37059  & B9 $^4$          & B9V  & B8V  &4.0$^{+1.0}_{-1.0}$&2.9 $\pm$ 0.3 &4.0$^{+1.0}_{-1.0}$ &2.3 $\pm$ 0.5 &3.0$^{+8.0}_{-0.0}$&2.8 $\pm$ 0.4 \\
32 & HD 37060  & A0n $^4$         & B8V  & -  &5.0$^{+0.0}_{-1.0}$&2.5 $\pm$ 0.3 &4.0$^{+7.0}_{-0.0}$ &2.7 $\pm$ 0.4 &4.0$^{+1.0}_{-1.0}$&2.4 $\pm$ 0.5 \\
33 & HD 37078  & A2V $^1$         & A0V+ & -  &5.0$^{+1.0}_{-1.0}$&2.1 $\pm$ 0.5 &4.0$^{+5.0}_{-1.0}$ &2.4 $\pm$ 0.7 &4.0$^{+2.0}_{-0.0}$&2.4 $\pm$ 0.5 \\
34 & HD 37115A & B6V(e) $^4$      & B6V*  & -  &0.6$^{+2.4}_{-0.1}$&5.0 $\pm$ 1.5 &0.9$^{+0.1}_{-0.1}$ &3.9 $\pm$ 0.2 &1.0$^{+0.0}_{-0.0}$&3.2 $\pm$ 0.1 \\
35 & HD 37129  & B3III/IV $^1$    & B3V* & -  &0.4$^{+0.3}_{-0.0}$&6.5 $\pm$ 1.5 &9.0$^{+2.0}_{-8.7}$ &6.3 $\pm$ 2.0 &5.0$^{+5.0}_{-4.7}$&7.1 $\pm$ 1.8 \\

36 & HD 37131  & B8V$^5$            & - & -  &7.0$^{+4.0}_{-5.0}$&3.9 $\pm$ 0.9 &3.0$^{+0.0}_{-1.0}$ &2.4 $\pm$ 0.6 &3.0$^{+0.0}_{-0.0}$&2.4 $\pm$ 0.1 \\

37 & HD 37150  & B2V $^3$         & B2V  & -  &0.4$^{+9.6}_{-0.2}$&8.6 $\pm$ 1.5 &0.4$^{+10.6}_{-0.2}$ &8.6 $\pm$ 1.5 &7.0$^{+4.0}_{-6.8}$&8.6 $\pm$ 0.7 \\
38 & HD 37174  & B9V $^{1,4}$ & B8V  & -  &4.0$^{+1.0}_{-1.0}$&2.7 $\pm$ 0.3 &4.0$^{+7.0}_{-1.0}$ &2.9 $\pm$ 0.4 &10.0$^{+0.0}_{-7.0}$&2.9 $\pm$ 0.6 \\
39 & HD 37334  & B3/5V $^1$       & B4V* & -  &0.4$^{+9.6}_{-0.1}$&6.7 $\pm$ 0.9 &0.6$^{+9.4}_{-0.2}$ &4.9 $\pm$ 1.6 &0.6$^{+9.4}_{-0.3}$&4.8 $\pm$ 1.9 \\
40 & HD 37356  & B3V $^1$         & B2V  & B2V  &0.3$^{+0.1}_{-0.1}$&5.9 $\pm$ 1.3 &0.3$^{+0.1}_{-0.1}$ &5.9 $\pm$ 1.2 &0.3$^{+0.1}_{-0.1}$&5.8 $\pm$ 1.1 \\

\end{tabular}

            }
    \label{tab:spec_class}    
    \end{table*}

    \begin{table*}
    \centering
    \renewcommand{\arraystretch}{1.3}
    \ContinuedFloat
    \caption{\textit{continued}}
    \resizebox{\textwidth}{!}{
        \begin{tabular}{ccccccccccc}\hline
Index & ID & \multicolumn{3}{c}{SpType} & \multicolumn{2}{c}{\textit{Gaia}} & \multicolumn{2}{c}{2MASS} & \multicolumn{2}{c}{WISE}\\
        &           & Ref          & IR     & Opt    & Age & Mass & Age  & Mass & Age & Mass \\ \hline
41 & HD 37428  & B9III/IV $^1$    & B8V* & -  &3.0$^{+1.0}_{-0.0}$&2.4 $\pm$ 0.2 &3.0$^{+0.0}_{-1.0}$ &2.4 $\pm$ 0.6 &3.0$^{+0.0}_{-1.0}$&2.5 $\pm$ 0.6 \\

42 & HD  37455 & A3Vb$^9$             & - & A0V+ &5.0$^{+1.0}_{-1.0}$&2.1 $\pm$ 0.3 &5.0$^{+2.0}_{-1.0}$ &2.1 $\pm$ 0.3 &5.0$^{+1.0}_{-1.0}$&2.1 $\pm$ 0.3 \\

43 & HD 37470  & B8/9III $^1$     & B8V* & -  &2.0$^{+0.0}_{-0.0}$&2.9 $\pm$ 0.2 &2.0$^{+0.0}_{-1.1}$ &2.9 $\pm$ 1.3 &2.0$^{+0.0}_{-0.0}$&2.9 $\pm$ 0.0 \\

44 & HD 37480  & A0V$^1$              & - & -  &4.0$^{+5.0}_{-1.0}$&2.7 $\pm$ 0.4 &5.0$^{+1.0}_{-0.0}$ &2.0 $\pm$ 0.2 &5.0$^{+1.0}_{-1.0}$&2.1 $\pm$ 0.3 \\

45 & HD 37545  & B8 $^{11}$       & B9V  & -  &4.0$^{+1.0}_{-1.0}$&2.6 $\pm$ 0.3 &4.0$^{+1.0}_{-1.0}$ &2.5 $\pm$ 0.4 &4.0$^{+1.0}_{-1.0}$&2.5 $\pm$ 0.4 \\
46 & HD 37547  & A0V $^1$         & B9V* & -  &5.0$^{+0.0}_{-2.0}$&2.7 $\pm$ 0.4 &5.0$^{+1.0}_{-1.0}$ &2.1 $\pm$ 0.5 &4.0$^{+2.0}_{-0.0}$&2.4 $\pm$ 0.6 \\
47 & HD 37687  & B7III $^2$       & B5V  & B5V* &0.7$^{+0.1}_{-0.1}$&4.2 $\pm$ 0.3 &0.7$^{+0.1}_{-0.1}$ &4.1 $\pm$ 0.4 &0.7$^{+0.1}_{-0.1}$&4.2 $\pm$ 0.3 \\
48 & HD 37700  & B8V $^1$         & B5V  & B6V  &7.0$^{+3.0}_{-6.1}$&4.3 $\pm$ 0.6 &10.0$^{+1.0}_{-9.0}$ &4.3 $\pm$ 1.3 &1.0$^{+10.0}_{-0.2}$&4.3 $\pm$ 1.3 \\

49 & HD  37745 & A0IV/V$^1$           & - & B9V &4.0$^{+1.0}_{-1.0}$&2.7 $\pm$ 0.7 &4.0$^{+1.0}_{-1.0}$ &2.3 $\pm$ 0.5 &4.0$^{+1.0}_{-1.0}$&2.3 $\pm$ 0.5 \\
50 & HD  37807 & B3/5IV $^1$          & - & B5V &0.9$^{+8.1}_{-0.2}$&4.8 $\pm$ 0.6 &0.9$^{+8.1}_{-0.0}$ &4.8 $\pm$ 0.5 &2.0$^{+1.0}_{-1.3}$&5.1 $\pm$ 0.9 \\

51 & HD 294275 & A1V $^{11}$      & A0V+ & A0V  &5.0$^{+1.0}_{-1.0}$&2.1 $\pm$ 0.5 &6.0$^{+1.0}_{-1.0}$ &1.9 $\pm$ 0.2 &6.0$^{+1.0}_{-2.0}$&1.9 $\pm$ 0.5 \\
52 & V1133 Ori & B9 $^5$          & B6V* & -  &6.0$^{+3.0}_{-5.1}$&4.3 $\pm$ 1.3 &2.0$^{+3.0}_{-1.0}$ &3.0 $\pm$ 1.5 &1.0$^{+8.0}_{-0.1}$&4.3 $\pm$ 1.3 \\
53 & V1148 Ori & ApSrEu $^1$      & -  & B8V* &1.0$^{+10.0}_{-0.2}$&4.5 $\pm$ 0.5 &2.0$^{+0.0}_{-1.0}$ &3.0 $\pm$ 1.3 &1.0$^{+10.0}_{-0.2}$&4.2 $\pm$ 1.2 \\\hline
\end{tabular}
            }
    \vspace{2mm}

\begin{minipage}{\textwidth}
\footnotesize
\textit{References:}
1 -- \cite{houk99};
2 -- \cite{abt04};
3 -- \cite{kyr22};
4 -- \cite{wolff04};
5 -- \cite{hsu13};
6 -- \cite{ner24};
7 -- \cite{kue24};
8 -- \cite{morgan66};
9 -- \cite{murphy20};
10 -- \cite{ramirez20};
11 -- \cite{cab19}.
\end{minipage}
    \end{table*}

    \begin{figure*}
        \centering
        \includegraphics[width=0.9\linewidth]{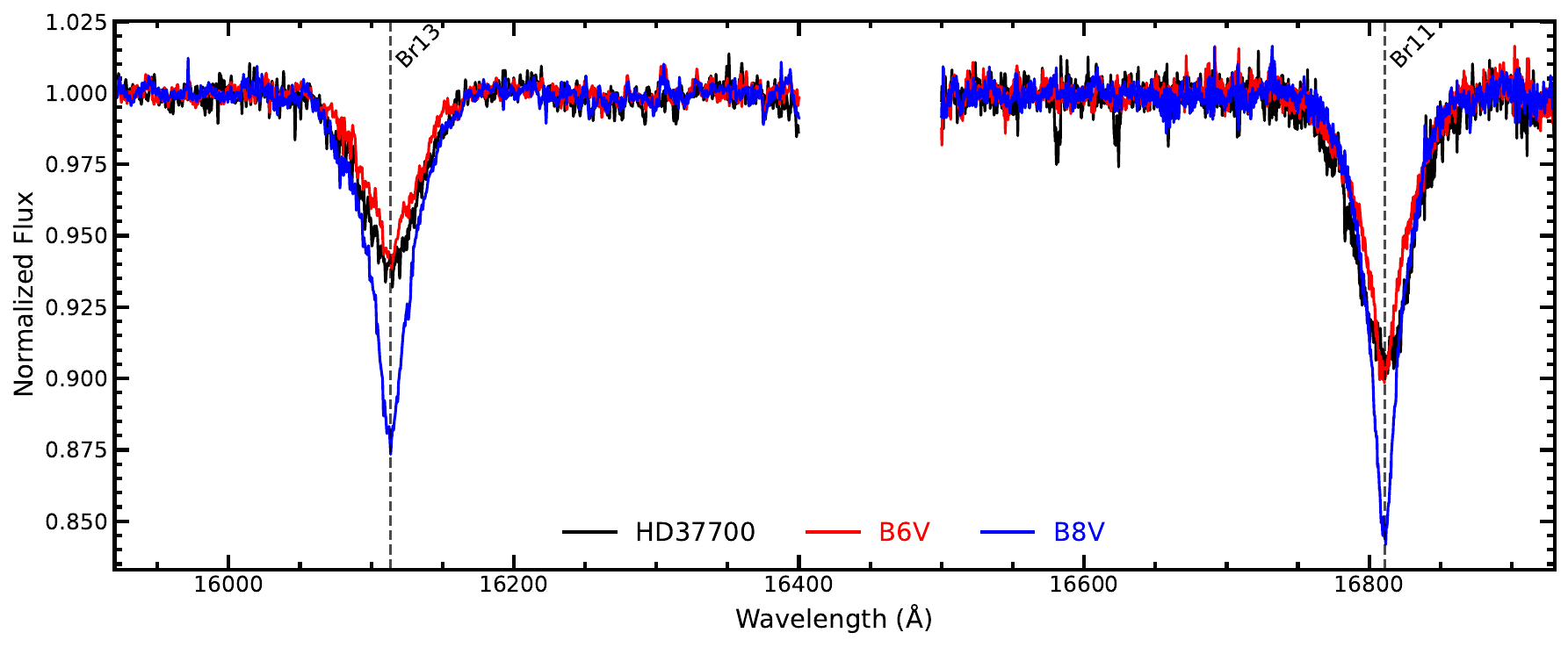}
        \includegraphics[width=0.9\linewidth]{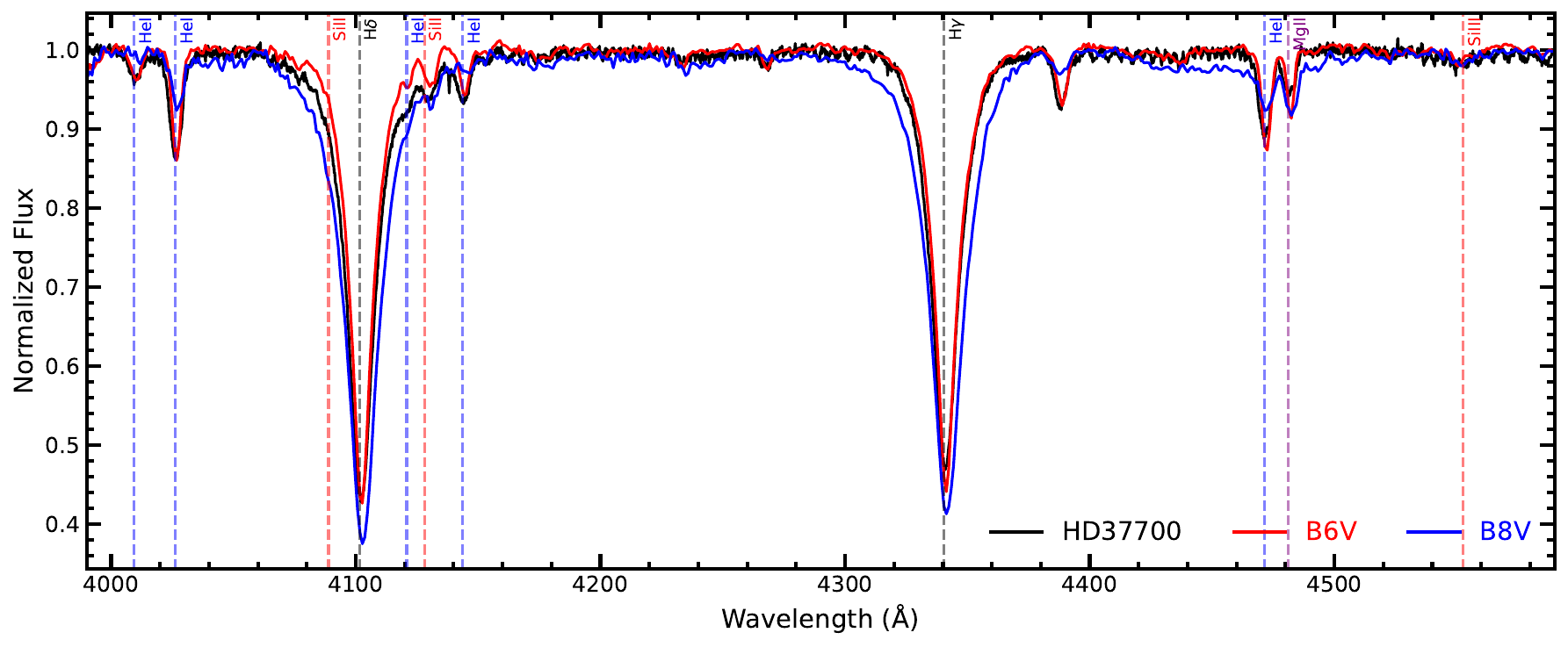}
        \caption{Spectra of HD\,37700 (\#48) (black solid line) and B6V (red) and B8V (blue) spectra template, in the IR and optical range in the top and bottom panel, respectively. The optical spectra of HD\,37700 and the template stars are from the LAMOST DR10 survey. The lines used for spectral classification, together with the Balmer series lines, are marked in both panels.}
        \label{fig:spec_class_likehood}
    \end{figure*}

\section{Spectral classification}
    \label{sec:spectral_classification}
    
    To estimate the spectral type from the IR spectra, we used the equivalent widths (EWs) of the Br11 (16811.111\,\AA) and Br13 (16113.714\,\AA) lines and their correlation with spectral type. However, instead of using the \citet{ramirez20} relation, we refined this correlation, as differences in the normalization process or updates in the APOGEE pipeline could lead to misclassification. 
    
    We extended the relation to include B2 stars, as this spectral type appears to be consistent with the linear trend \citep[see Fig.\,5 of][]{ramirez20}. From their list, we selected three stars of each spectral subtype (B2--A0) and used them as template stars to refine our correlation, as expressed in equation.\,\ref{Eq_spt_relation}:
    
    \begin{equation}
        \label{Eq_spt_relation}
        {\textrm SpType} = 0.798\,\textrm{EW}[{\rm Br11} + {\rm Br13}] + 6.961 (\pm 1)
    \end{equation}
    
    From this relation, stars later than A0 cannot be reliably classified, since the EWs of Br11 and Br13 begin to decrease for these spectral types, causing potential misclassification as late B-type stars. For this reason, we did not apply this method to stars identified as later than A0 in the optical spectra or in the literature.

    For the optical spectra, we identified the characteristic lines of B stars and measured their EW ratios. For early B stars (B0--B2), we used the ratios Si\,\textsc{iii} 4552\,\AA\ / Si\,\textsc{iv} 4089\,\AA\ and Si\,\textsc{iii} 4552\,\AA\ / He\,\textsc{i} (4009, 4026, 4121, 4144\,\AA), while for later types stars (B3--A0), the He\,\textsc{i} 4471\,\AA\ / Mg\,\textsc{ii} 4481\,\AA\ ratio, along additional ratios between Si\,\textsc{ii} (4128--4130\,\AA) and He\,\textsc{i} (4009, 4026, 4121, 4144\,\AA) lines were used \citep{graycorbally09,sota11,ramirez20,ner24}. The EW ratios obtained for each star were then compared with the ratios of the template stars, and the spectral type was assigned based on the best match between them.
    
    For both classification methods, each spectrum was carefully compared with the template spectra of the corresponding spectral type to verify the match. The same stars used to derive the IR Brackett-line relation were also adopted as templates for both IR and optical ranges. This approach helps to avoid misclassification due to peculiarities, such as weak He lines.
    
    To assign the luminosity class for the star in our sample, we searched in the optical spectra for diagnostic lines of class I-III stars, such as N\,\textsc{ii} and Si\,\textsc{ii} lines \citep{graycorbally09}. We also examined the intensity and wings profiles of the Balmer lines, which are sensitive to surface gravity \citep{ner24}. For stars that do not have an optical spectrum available, the IR classification method is particularly sensitive to luminosity classification for stars later than B5, but not for earlier-type stars \citep{ramirez20}. Therefore, for stars of spectral type B5 or earlier without an optical spectrum -- stars \#8, 20, 37, and 39 -- we searched in the literature for their log\,\textit{g} values to confirm their luminosity class. According to \cite{Sprague22}, these stars have log\,\textit{g}$\simeq4$, which are typical values for MS stars, thus excluding the possible classification as class I-III stars. 

    \begin{figure}
        \centering
        \includegraphics[width=\linewidth]{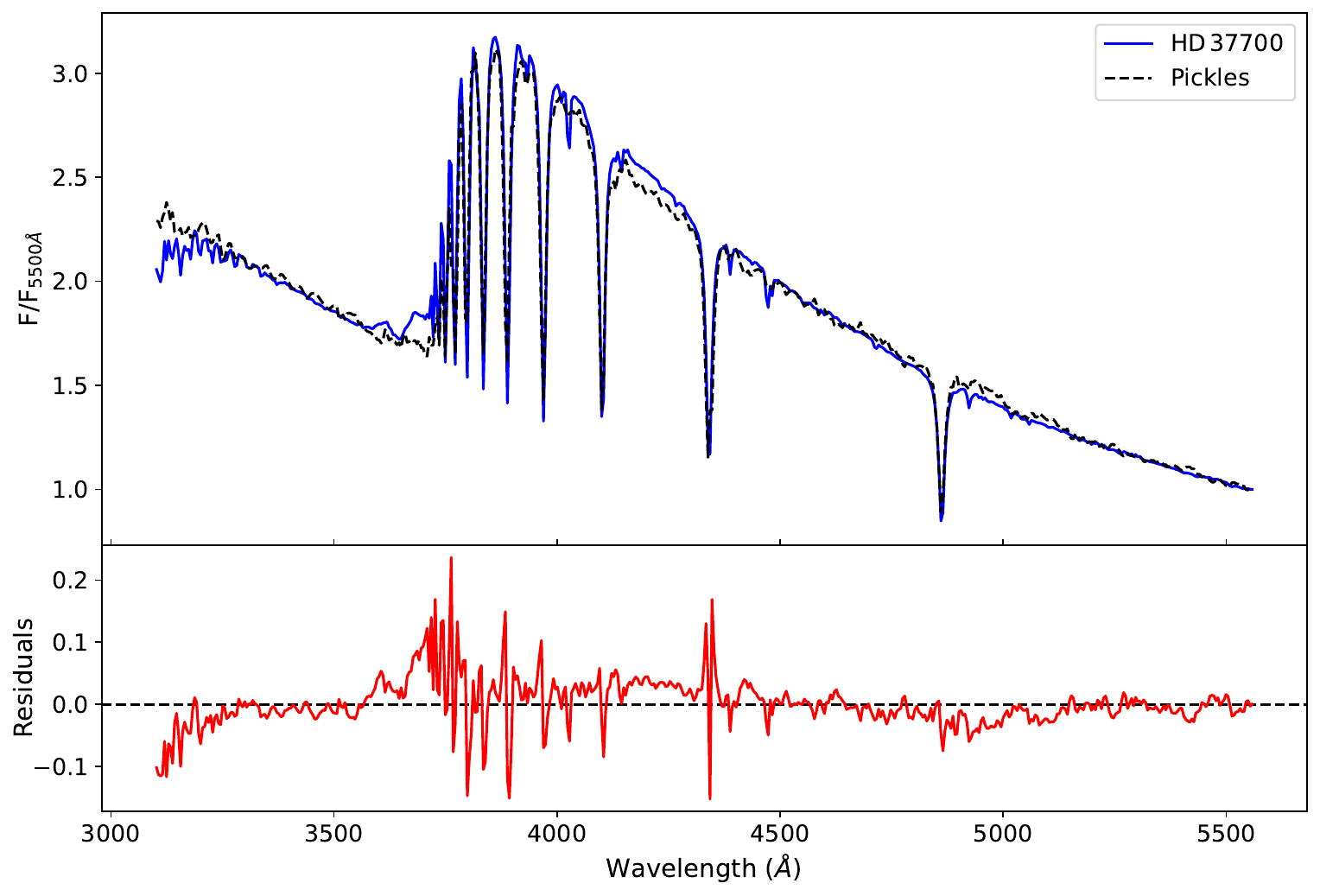}
        \caption{Best fit of the observed optical spectra of HD\,37700 (\#48) (blue solid line) corrected by $E(B-V)=0.07$\,mag and the Pickles (1998) template spectra (black dashed line) in the top panel. In the bottom panel, the relative residual difference (red solid line) is shown, with the zero dashed line shown for visual guidance.}
        \label{fig:reddening_spec_pickles}
    \end{figure}
    
    The columns ``IR'' and ``Opt'' of Table\,\ref{tab:spec_class} list the spectral types derived from the IR and optical spectra, respectively, together with the luminosity classification for the entire sample. Fig.\,\ref{fig:spec_class_likehood} shows the spectra of HD\,37700 (\#48) compared to the B6V (this work) and B8V \citep{houk99} templates in the IR (top) and optical (bottom) ranges. The lines used in the classification process, together with the Balmer series, are provided to illustrate the difference in EW among stars of different spectral types. It is noticible that the spectra from a B6V star provide a better match to HD\,37700 than a B8V.
    
    The classification from IR and optical spectra, as well as those from the literature, which we obtained from the \cite{Skiff09} database, generally agree within one spectral subtype (see Table\,\ref{tab:spec_class}). In some cases, the IR classification yielded a spectral type earlier than B2, caused by faint Br11/Br13 lines or poorly defined wings; in these situations, the classification was based solely on visual comparison with templates (cases indicated by an ``*'' in Table\,\ref{tab:spec_class}). Despite the qualitative nature of this procedure, the classifications proved coherent with those reported in the literature.
        
\section{Age, Mass and Reddening Determination}
    \label{sec:age_mass_estimation}
    
    The study of the ON region allows the application of strong constraints to the isochrone fitting due to previously established parameters, such as the distance. For the distance modulus correction, we adopted two different values depending on stellar age: older populations ($\tau \geq 5$ Myr) are located closer to us, with $(m - M)_0 = 7.95$ ($\sim$385\,pc), while younger populations ($\tau < 5$ Myr) have $(m - M)_0 = 8.1$ ($\sim$410\,pc) \citep[e.g.][]{maia10,alvesbouy12,bouy14,braz22}.
    
    Using the metallicity derived from the ASPCAP pipeline (Holtzman et al. in prep.), we determined an average value for the crossmatch between the APOGEE and the sources from the crossmatch of \textit{Gaia}, 2MASS and WISE surveys with $2 \leq \varpi \leq 3$\,mas, which results in 1032 stars. From these stars, we derived an average value of [M/H] $\simeq -0.2$, which was adopted as a second constraint for the isochrone fitting procedure. Since the most distant stars in the sample (\#40, 41, 47, and 52), with $D\geq 425$\,pc, are located at the edge of the ON (see Fig.\,\ref{fig:skymap_crossmatch}) and the stars along its line of sight (\#11, 17, 20, 21, 22, 32) are in the foreground, with $D\leq 380$\,pc, we assumed a uniform reddening of $E(B-V) = 0.1$. 
    
    This value is consistent with the observed distribution of this population and agrees with the reddening derived for the nearby open clusters in the ON region, such as $\sigma$ Orionis ($E(B-V)=0.05$), NGC\,1977 ($E(B-V)=0.11$), NGC\,1980 ($E(B-V)=0.06$), NGC\,1981 ($E(B-V)=0.07$) and the foreground population of Trapezium Cluster ($E(B-V)=0.08$) \citep{maia10,Dias21}. Moreover, the adopted reddening is also consistent with the individual $E(B-V)$ values for each sources in our sample available in the literature \citep{ZorecBriot91,Wegner02,Hernandez05,rom21}.

    To confirm whether this reddening value mentioned above is consistent with reddening maps, we used the coordinates and distances of the 53 stars in our sample to compute the $E(B-V)$ by using the maps of \cite{Vergely22} and \cite{Edenhofer24}. The maps of \cite{Vergely22} are available at three resolutions: 10, 25, and 50\,pc. The average $E(B-V)$ for those three resolutions is 0.178, 0.129, and 0.096\,mag, respectively. It should be noted that there is a compromise between resolution and the number of objects used to constrain the extinction. Nevertheless, across the three resolution ranges, $E(B-V)$ varies from approximately 0.096 to 0.178, with an intermediate value of 0.129, which is consistent with the ranges of $E(B-V)$ considered in the present work. In the case of \cite{Edenhofer24}, the average $E(B-V)$ is 0.083, still consistent with the values considered in the present paper.

    \begin{table}
        \centering
        \caption{Reddening estimated by comparing optical spectra to the Pickles (1998) template spectra. The first column shows the stars ID and the second column the $E(B-V)$ values.}
        \begin{tabular}{c|c}\hline
                ID     & E(B-V) \\\hline
            HD\,36540  (\#5)   & 0.18 \\
            HD\,36629  (\#6)   & 0.26 \\
            HD\,36918  (\#13)  & 0.07 \\
            HD\,36919  (\#14)  & 0.08 \\
            HD\,37017  (\#25)  & 0.11 \\
            HD\,37058  (\#30)  & 0.02 \\
            HD\,37687  (\#47)  & 0.26 \\
            HD\,37700  (\#48)  & 0.07 \\
            HD\,294275 (\#51)  & 0.08 \\\hline 
        \end{tabular}
        \label{tab:reddning_spec_pickles}
    \end{table}

    An alternative way to estimate reddening is to match the observed optical spectra to template spectra from \cite{Pickles98} empirical library, adjusting $E(B-V)$ using an extinction law by \cite{Cardelli89} with R$_{V} =$ 3.1. Before applying the match procedure, the observed spectra were binned to approximate the resolution of the template spectra (5A) and both normalized to unity at a featureless continuum wavelength.

    \begin{figure*}
        \centering
        \includegraphics[width=\textwidth]{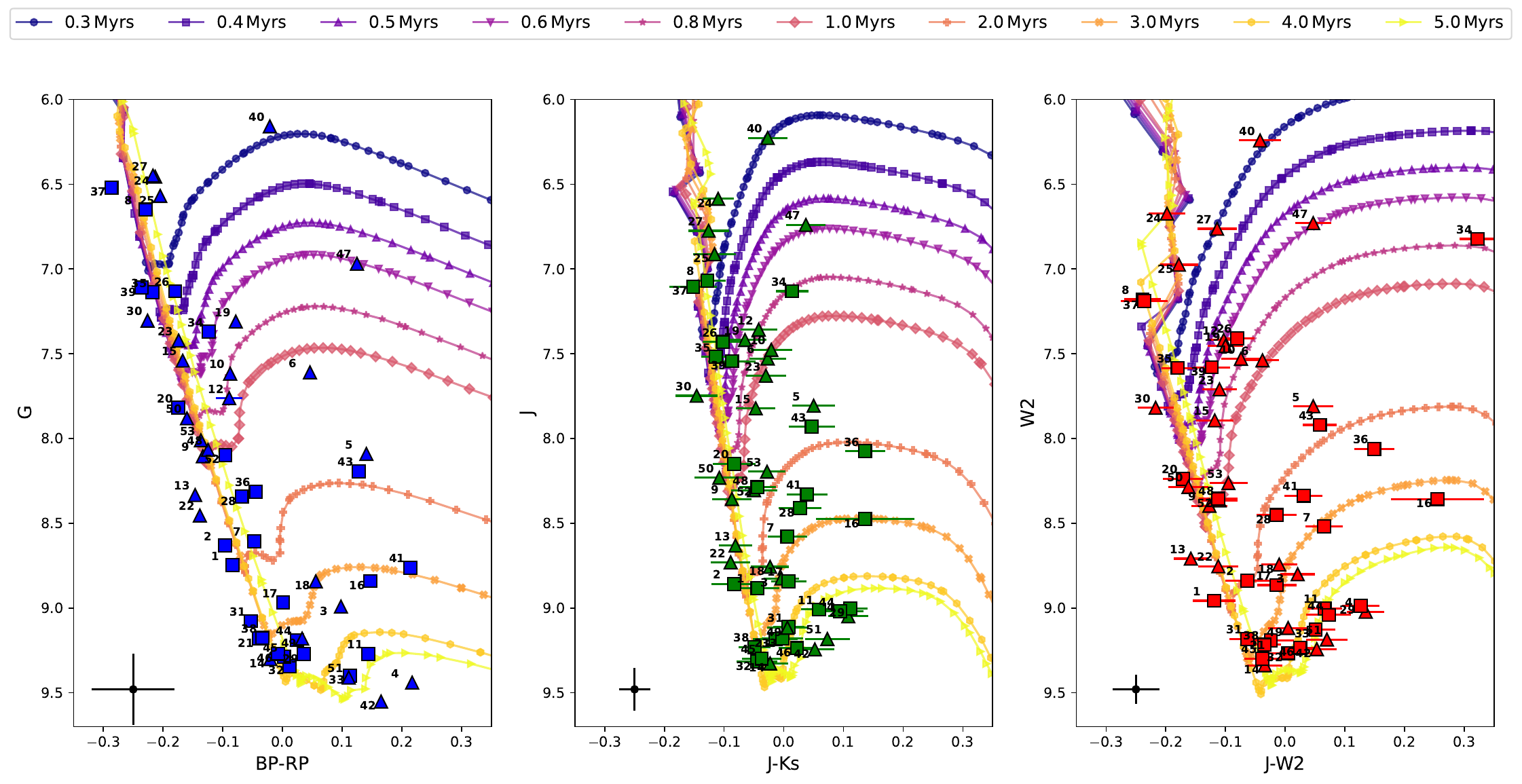}
        \caption{CMDs of \textit{Gaia} (left panel), 2MASS (middle panel), and WISE (right panel) surveys with propagated photometric errors for each star. The squares represent stars with only IR spectra, while the triangles represent objects with both optical and IR spectra. Isochrones of different ages are plotted with distinct colours and symbols, as indicated in the legend, spanning from 0.3 Myr (darker colours) to 5.0 Myr (lighter colours). The uncertainty in distance modulus of $(m - M)_0 = 0.1$, corresponding to 15\,pc, and in reddening of E(B-V) = 0.05\,mag, adopted to derive the mass and age uncertainties of the sample are shown at the bottom left of each panel.}
        \label{fig:cmds_isocs}
    \end{figure*}

    The best $E(B-V)$ was obtained by the lowest $\chi^2$ achieved for the flux difference between the observed spectrum and the template of the closest spectral type. In the process, the $E(B-V)$ value was varied from -0.2 to 1.0 in steps of 0.01, the lower limit set to allow the search for values close to zero reddening.

    The spectral match for HD\,37700 (\#48) with the corresponding residuals, i.e., (observed-template)/template, is shown in Fig.\,\ref{fig:reddening_spec_pickles}. In this case, the determined $E(B-V)$ is equal to 0.07\,mag. The reddenings derived in this way for our sample are listed in Table\,\ref{tab:reddning_spec_pickles}. The cases where the $E(B-V)$ is discrepant from the adopted value, whether by our determination or from the literature, are commented in Sect.\,\ref{Sect:special_cases}.

    All spectra with either $E(B-V)$ negative (which indicates that the observed spectrum needs to be reddened rather than underreddened) or greater than 0.5 were discarded from our estimates. Possible reasons for those values include non-peculiar spectra, presence of circunstellar material, incorrect calibration fluxes, and a unique R$_V$ value (in our case, 3.1), among others.
    
    To estimate stellar ages and masses, we constructed a grid of PARSEC isochrones v3.8 \citep{bre12} spanning 0.1--1.0\,Myr in steps of 0.1\,Myr, and 1--11\,Myr in steps of 1\,Myr. For each isochrone in the grid, we computed the $\chi^2$ independently for the three CMDs -- \textit{Gaia}, 2MASS and WISE -- in the photometric space. This was done by comparing the stellar colour and magnitude with each point along the isochrones, accounting for the corresponding observational uncertainties, to identify the closest match between each star and the model. The adopted age and mass correspond to the point in the isochrone that minimises the $\chi^2$ value.

    \begin{figure*}
        \centering
        \includegraphics[scale=0.6]{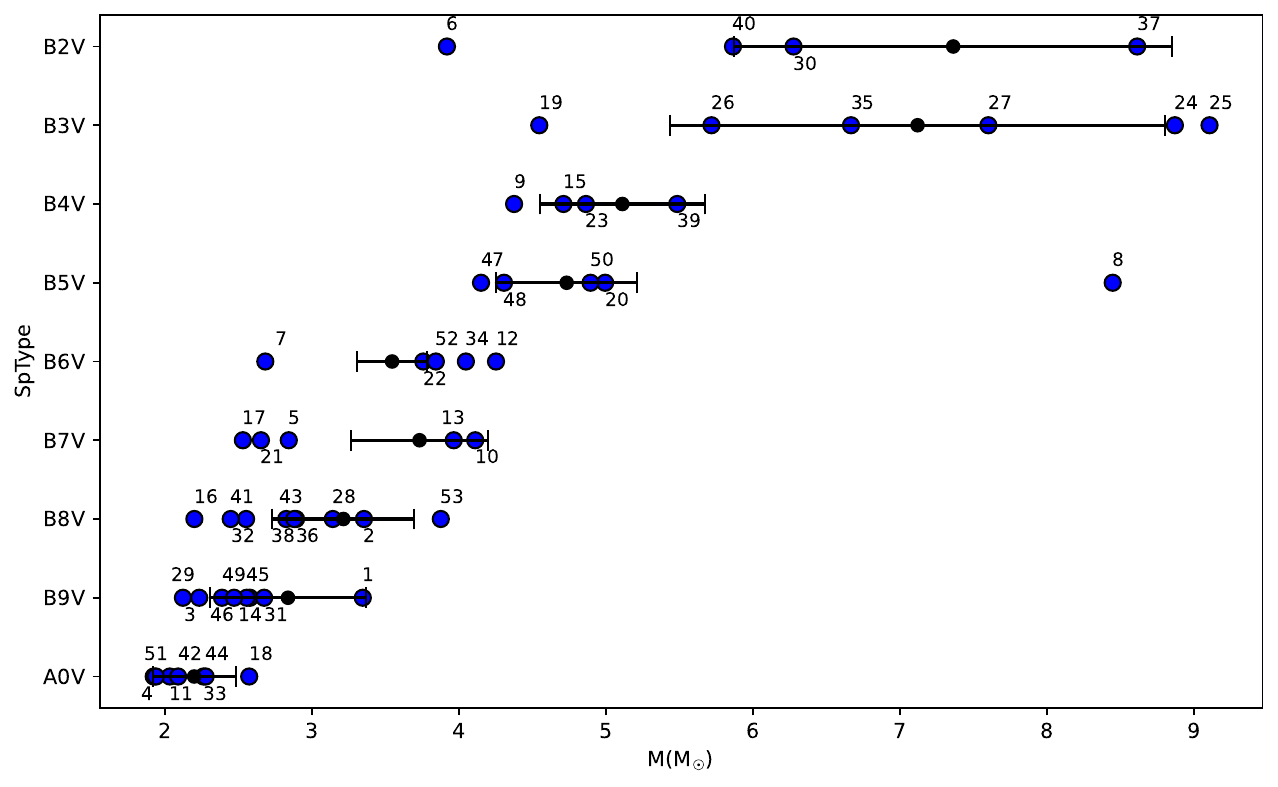}
        \caption{Mass estimate for each star in our sample (blue circles) and for the eclipsing binary stars from Eker et al. (2018) with their corresponding uncertainties (black circles with error bars), derived from Gaussian fit. Stars later than A0 were grouped with the A0 subtype, as the mass difference among these spectral types is not significant.}
        \label{fig:mass_spt_relation}
    \end{figure*}

    Uncertainties were estimated by varying the distance modulus and reddening of the isochrones by $\pm$0.1 and $\pm$0.05, respectively. For each combination of these parameters, we computed the $\chi^2$ value and selected the solutions that minimised it. Then, the difference between the most extreme values of mass and age obtained in these variations and the best-fit values was adopted as the uncertainty. For the age estimates, the uncertainties were separated into upper and lower limits. Cases in which the age uncertainty is reported as zero indicate that the estimated age remained unchanged.

    Figure~\ref{fig:cmds_isocs} shows the CMDs for the three surveys: \textit{Gaia} (G vs.\ $G_{BP} - G_{RP}$), 2MASS (J vs.\ $J - K_s$), and WISE (W2 vs.\ $J - W2$). Isochrones of different ages are overplotted to illustrate the separation between stellar populations of distinct ages and evolutionary stages. The resulting mass and age estimates, with the uncertainties, are listed in Table~\ref{tab:spec_class}.

    The assumption of a uniform reddening across a star-forming region can introduce systematic biases in the isochrone fitting. Nevertheless, the stellar population can generally be modeled with isochrones that share the same parameters -- except for age in this case -- since these stars likely share a common origin. Although the effect of the reddening differs among the three CMDs, the estimated masses and ages remain consistent, within the uncertainties, across all of them, indicating that the adopted reddening is appropriate to this population.

    \begin{table}
        \centering
        \caption{Mass threshold per spectral type determined from the dynamical mass measurements of eclipsing binary stars.}
        \begin{tabular}{cc}\hline
        SpType & Mass (M$_\odot$)\\\hline
        B2V & 7.4 $\pm$ 1.5 \\
        B3V & 7.1 $\pm$ 1.7 \\
        B4V & 5.1 $\pm$ 0.6 \\
        B5V & 4.7 $\pm$ 0.5 \\
        B6V & 3.5 $\pm$ 0.2 \\
        B7V & 3.7 $\pm$ 0.5 \\
        B8V & 3.2 $\pm$ 0.5 \\
        B9V & 2.8 $\pm$ 0.5 \\
        A0V & 2.2 $\pm$ 0.3 \\\hline
        \end{tabular}
        \label{tab:mass_spt_relation}
    \end{table}

    To validate our mass estimates and ensure that the appropriate isochrones were applied, we compared our results with the dynamical masses of eclipsing binaries from \citet{eker18}, using them to define the expected mass range for each spectral type. Gaussian profiles were fitted to the mass distribution of each subtype to determine the mean and corresponding dispersion of $1\sigma$ (Table~\ref{tab:mass_spt_relation}). Most stars agree with the expected mass values within one spectral subtype, as can be seen in Fig.\,\ref{fig:mass_spt_relation}, where the blue circles indicate the mass of each star in our sample, while black circles with error bars represent the gaussian fit from eclipsing binary stars. 
    
    Stars \#6, 7, and 8 show a significant discrepancy between their estimated mass and the assigned spectral type. Due to its peculiarities, these cases are discussed in Sect.\,\ref{Sect:spectral_mass_outliers}. Although some stars, such as \#17, 19, 21, and 25, appear to deviate from the general trend, their uncertainties -- omitted from the plot for better readability -- make them consistent with their spectral type or at least within one subtype. This agreement between our mass estimates and those from empirical constraints supports the reliability of both our spectral classifications and age determinations, which is particularly important given that different sets of isochrones can yield different masses.

\section{PMS Candidates}
    \label{sec:pms_candidates}

    Luminosity classes I--III were previously assigned to some stars in our sample. However, neither these objects nor any other stars in the sample show spectral features indicative of this classification. Instead, their properties are more consistent and similar to those of class V stars. This evidence suggests that the objects located along the PMS tracks are young stellar objects, contracting towards the MS rather than evolved stars.

    This is the case of the stars \#3, 4, 5, 11, 16, 29, 36, 41, 42, 43, 44, 47, and 49, which are consistently found on the PMS track in all three CMDs (Fig.~\ref{fig:cmds_isocs}), making them the most secure PMS candidates, being classified as robust PMS. Stars \#25, 26, 39 and 40 lie on the PMS track in the WISE CMD but are close to the MS in the 2MASS and \textit{Gaia} CMDs; based on additional characteristics discussed in Sect.~\ref{Sect:special_cases}, they are being classified as peculiar PMS candidates.

    Accordingly, in Fig.\,\ref{fig:cmds_isocs}, most stars appear on the MS in the optical CMD but lie along the PMS track in the IR CMDs, within the uncertainties. This is the case for stars \#7, 10, 12, 17, 19, 23, 28, 33, 34, 46, 51, 52, and 53, which we classify as likely PMS. Due to its distinct reddening and extinction law, the star \#6 is also included in the likely PMS group (see Sect.\,\ref{Sect:spectral_mass_outliers}). Infrared CMDs are particularly valuable for identifying young stellar objects, as they are less affected by reddening than optical CMDs. In addition, young stars often display IR excess due to circumstellar material, which may account for their distinct positions in the CMDs.
    
    For the remaining objects, the stars \#1, 9, 20, 24, 30, 35, and 48 lie on or very close to the MS and exhibit the largest discrepancies in age estimates across the three CMDs (see Fig.\,\ref{fig:cmds_isocs} and Table~\ref{tab:spec_class}). For these stars, precise ages could not be determined due to strong degeneracy in the result -- that is, within the uncertainties, their positions are consistent with multiple evolutionary tracks. Furthermore, stars \#2, 8, 13, 14, 15, 18, 21, 22, 27, 31, 32, 37, 38, 45 and 50 yield compatible ages in at least two CMDs, but their position and uncertainties in the CMD suggest that they should also be considered MS stars.

    \subsection{Ages Distribution}
    
        The stellar age is a key parameter for unveiling the evolutionary history of a star-forming region, as it allows the reconstruction of its initial formation scenario \citep[e.g.][]{bouy14,kounkel18,grossschedl21}. Moreover, stellar age is a crucial parameter for understanding the broader context of the Milky Way disc formation and evolution \citep{Amores17}. 
        
\begin{figure}
            \centering
            \includegraphics[width=\linewidth]{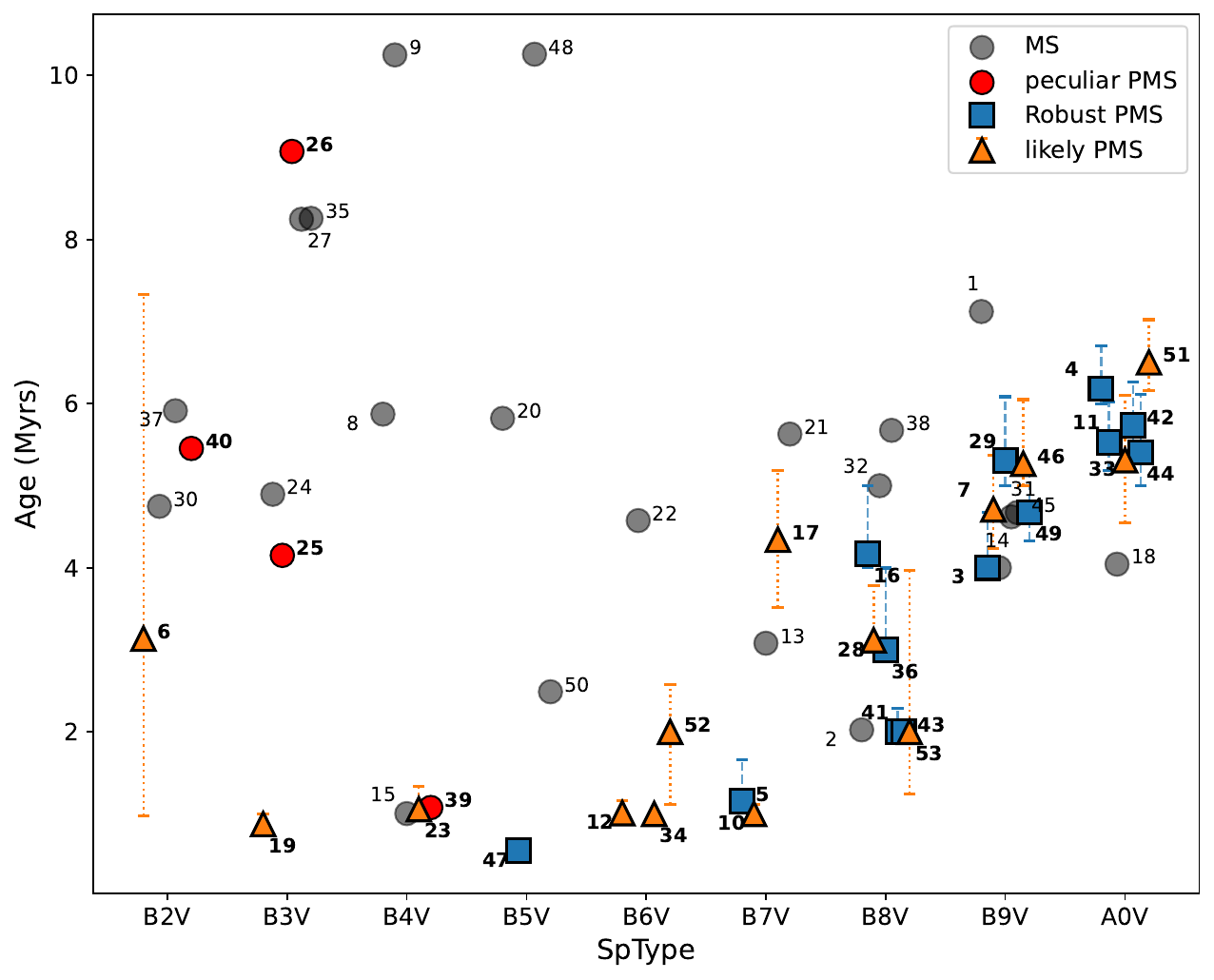}
            \caption{Age-spectral type diagram of B and early-A type candidates surrounding the Orion Nebula region. Stellar ages were derived from a Monte Carlo simulation, with the median age adopted as the representative age and the 16$^{\textrm{th}}$ and 84$^{\textrm{th}}$ percentiles as the lower and upper uncertainties, respectively. The blue squares, green triangles, and red circles represent the robust, likely, and peculiar PMS candidate groups, respectively, with their IDs shown in bold capitals. The main-sequence (MS) stars are represented by the grey circles.}
            \label{fig:age_SpT}
        \end{figure}

        To investigate the age distribution within our sample, we performed a Monte Carlo simulation in which 10000 age values were generated from a split-normal distribution centred on the estimated age and parameterized by the corresponding upper and lower uncertainties, for each catalogue. The ages of the three catalogues were then combined to derive a single age estimate for each star. We adopted the median of the resulting distribution as the representative age, while the 16$^{\textrm{th}}$ and 84$^{\textrm{th}}$ percentiles were taken as the lower and upper uncertainties, respectively. The resulting age distribution is shown in Fig.\,\ref{fig:age_SpT}. For clarity, the robust, likely, peculiar PMS candidate and the MS groups, as defined in Sect.\,\ref{sec:pms_candidates}, are represented by different symbols.

        Within the limitations of our sample, among the PMS candidates (robust and likely), the fitted ages show an apparent distribution in which the earlier-type stars (B3V--B5V) tend to occupy younger age bins, while the later-type stars are preferentially associated with older ages. In contrast, the stars classified as MS objects span a broad range of ages across different spectral types, most likely because of the larger uncertainties in their age estimates (see Table.\,\ref{tab:spec_class}).

        We do not interpret this apparent pattern as evidence for an intrinsic age–mass relation, since owing to the crossmatch among the three catalogues and the adopted selection criteria, our sample is incomplete for highly extincted objects and may therefore miss very young stars exhibiting disc signatures. For example, HD\,36917, a young Herbig B9 star showing H$\alpha$ emission and deeply embedded in the ON, is excluded from our sample because of its large colour index, $J-W2\simeq 3$\,mag.

        Nevertheless, the presence of B stars in different evolutionary phases -- both PMS and MS -- with consistent ages spanning from less than one to a few million years is consistent with multiple star formation episodes occurring, likely triggered by distinct events. The younger early-type stars may have formed through feedback from older nearby massive populations \citep{maia10,bouy14}, whereas the older population may instead be a consequence of the large-scale events that shaped the Orion complex region \citep{kounkel18,grossschedl21}.
    
        Regarding the formation of early-type stars, some of the youngest PMS and almost all of the runaways candidates are located in relatively gas-free regions, farther from the ON, such as the stars \#6, 28, 40, and 52 (see Fig.\,\ref{fig:skymap_crossmatch}). Given their high masses, these stars may have played a significant role in clearing the gas in their formation environments. Nevertheless, early-type stars are typically not formed in isolation, but rather in clustered environments \citep{motte18}; therefore, it cannot be ruled out that these stars may have been ejected from their progenitor molecular cloud. A trajectory reconstruction of the original birthplaces of these B stars deserves further investigation to provide more insights about the star formation history of ON.
    
\section{Peculiar Characteristics}
    \label{Sect:special_cases}

    To further characterise our sample and examine cases that warrant particular attention due to peculiar characteristics, such as spectroscopic variability, nebular emission, or narrow spectral lines, we summarise and discuss specific properties in what follows. Additionally, when necessary, the mass and age were re-estimated by using the same isochrone set, but adopting the revised stars distance and/or reddening values to corrected them, as these peculiar features may have led to spectral misclassification or to overestimation of the mass and age.

    \subsection{Spectral Features}
    
        Stars \#22, 23, and 31 exhibit clear signatures of nebular emission lines in their spectra, such as O\,\textsc{iii}, N\,\textsc{ii} and Si\,\textsc{ii} lines. The presence of nebular emission can contaminate the Balmer series lines either strongly, as in the case of \#22, which shows emission in all the Balmer lines, or more weakly in the other two cases, where emission is primarily observed in H$\alpha$. The contamination makes it difficult to determine whether the H$\alpha$ emission originates from the star itself or from the surrounding nebula. Six stars in our sample -- \#6, 19, 30, 40, 47, and 50 -- display notably narrow spectral lines, which is consistent with their classification as slow rotator stars, as previously reported in the literature for some of these objects \citep{rom21,ner24}.
        
        As expected, a small fraction of our sample consists of Herbig stars. Among the robust, likely and peculiar PMS groups, 12 out of 32 were previously classified as Herbig stars, while for the MS group, 6 out of 22 stars have been identified as Herbig stars \citep{Hernandez05}. Except for \#50, the Herbig stars labelled as MS stars in this work -- \#2, 8, 24, 27, and 48 -- appear to lie on the PMS tracks or very close to the MS in at least one CMD (see Fig.\,\ref{fig:cmds_isocs}). However, their photometric uncertainties prevent a definitive confirmation of their PMS status. In contrast, the robust, likely and peculiar PMS Herbig stars -- \#3, 4, 5, 7, 25, 28, 29, 39, 40, 44, 49, and 53 -- are clearly located in the PMS phase.
    
        Notably, among all previously identified Herbig stars in our sample with available optical spectra, only \#25 and 53 exhibit a very weak H$\alpha$ emission component. The low resolution of LAMOST, from which most of the optical spectra were obtained, may be insufficient to resolve weak emission components in the remaining objects, particularly if they are similar to \#25 and 53, which were observed with the high-resolution GHOST and FEROS spectrograph, respectively. Alternatively, the absence of H$\alpha$ emissions may be explained by the rapid disc evolution in early-type stars. Given that these objects show no clear spectroscopic signature of circumstellar discs and have ages of a few Myrs, significantly longer than the typical disc lifetime of $\sim$ 0.1\,Myrs, their discs may have already dissipated \citep{AlonsoAlbi09}.

        Lastly, star \#34 exhibits emission lines throughout the entire IR spectrum. Since these emission lines are contaminating the Brackett lines, its spectral type could not be derived in this work. However, previous studies have classified it as B6Ve PMS star \citep{wolff04,cho15,Campbell23}, which is consistent with our isochrone-based mass estimate within one subtype (tables~\ref{tab:spec_class} and \ref{tab:mass_spt_relation}). We therefore adopt the B6V spectral type classification for this star.

    \subsection{Spectroscopic Variability}
    \label{Sect:spec_var}

    Stars exhibiting spectroscopic variability represent particularly interesting science cases for follow-up investigations. Such variabilities may be connected to the activity of the stars with their circumstellar material, magnetic activity, or binarity and therefore provide important diagnostics for understanding the evolution of early-type stars. The spectra of these stars are provided in the supplementary online material.

        \subsubsection{HD 36827 (\#8)}
    
            This star exhibits spectroscopic variability in its IR spectra. It hosts a very low-mass companion ($M\approx0.04M_\odot$) \citep{ker19}, and it has also been classified as a young stellar object candidate \citep{mar23}. These findings suggest that the observed spectroscopic variability is more likely driven by interaction with its circumstellar environment rather than binarity, as this star was also classified as a Herbig star \citep{Hernandez05}. The Br11/Br13 lines relation indicates a spectral type of B5V, but considering its relatively high mass ($M = 8.1 \pm 2.1\,M_\odot$) and the close resemblance to the best-matched template spectra, a classification as B4V or earlier appears more plausible.

        \subsubsection{HD 37025 (\#26)}
    
            This star shows weak Br13 profiles and strong spectroscopic variability in the IR spectra. Consequently, its classification was not possible using either the Brackett-line relation or template comparison, as variability prevents a reliable match. Its mass estimate of approximately $M \approx 5\,M_\odot$ suggests a star of spectral type earlier than B4V, which is consistent with the literature (see Table\,\ref{tab:spec_class}). Given that the strong spectroscopic variability is a common characteristic of PMS objects, this star is very likely a PMS candidate.

        \subsubsection{HD 37334 (\#39)}
    
            This star exhibits spectroscopic variability in the IR spectra. It was classified as a Herbig star and has a close companion at 2.5\,au with a mass of $M \approx 0.4\,M_\odot$ \citep{ker19}. Therefore, further investigation is required to confirm whether the variability is due to its close companion or something else.  
    
            The Br13 line is weak and shows absorption components, likely leading to misclassification by the Brackett-lines method. Template-based classification was also hindered by numerous metallic lines, probably originating from the later-type companion. Although no definitive spectral type could be assigned with our methodology, photometry and mass estimates (see Fig.\,\ref{fig:cmds_isocs} and Table\,\ref{tab:spec_class}) are consistent with an early-type star, likely B4V.

        \subsubsection{HD 37428 (\#41)}
    
            This star displays spectroscopic variability in the IR spectra. A magnetic field has been detected \citep{rum23}, which could explain the variability, although further confirmation is still required. As the most distant object in the sample ($\sim$500\,pc), its estimated age and mass at this distance are $2^{+0.0}_{-1.0}$\,Myrs and $M = 2.8 \pm 1.0\,M_\odot$, respectively, in good agreement with the previous estimated values, within the uncertainties.

        \subsubsection{V1133 Ori (\#52)}
    
            This star displays spectroscopic variability on its spectra and has been identified as a chemically peculiar magnetic object. Its radial velocity variations led to its classification as a probable binary star \citep{rom21}. Adopting its small reddening of E(B-V) = 0.02 and distance of $\sim$470\,pc in the isochrone fitting procedure, the estimated age and mass are $1^{+1.0}_{-0.1}$\,Myrs and $4.0\,\pm\,1.1\,M_\odot$, respectively, which is consistent with its B6V spectral type classification. Further investigation is required to confirm the origin of the spectroscopic variability, which may be caused either by its binarity or by the interaction of its magnetic field with the surrounding environment.

    \subsection{Stars With Discrepant Masses}
    \label{Sect:spectral_mass_outliers}

        Since the discrepancies that place some stars outside the expected mass range for their assigned spectral types are often linked to distances inconsistent with the adopted $(m-M)_0$ or reddening values, we recalculated the masses and ages of these outliers using \textit{Gaia} distance and reddening estimates from our method (Sect.\,\ref{sec:age_mass_estimation}) or from the literature. This procedure was also applied to stars with consistent mass estimates but exhibiting significantly higher reddening than the value adopted in the isochrone fitting. These stars are listed below.
    
        \subsubsection{HD 36540 (\#5)}

            This object is a chemically peculiar Herbig star located at a distance of 420\,pc, with reddening of $E(B-V)=0.18$ based on our determination, which is in good agreement with the literature \citep{Hernandez05,rom21}. Adopting these values for reddening and distance, we estimate a mass of $4.1\pm1.2\,M_\odot$ and an age of $1^{+1.0}_{-0.1}$\,Myrs, which are more consistent with its B7V spectral classification (see Fig.\,\ref{fig:mass_spt_relation}).

        \subsubsection{HD 36629 (\#6)}
        
            This star is a well-known prototype of a B2V star \citep{ner24}. The extinction law in its direction is $R_v$ = 3.36, its reddening is E(B-V) = 0.26 from our determination, and it lies at a distance of 423\,pc \citep{Wegner02}. Adopting these parameters in the isochrone fitting we now estimate an age of $0.7^{+4.0}_{-0.4}$\,Myrs and mass of $M = 7.1 \pm 1.9\,M_\odot$, with the upper large uncertainty is due to its proximity to the zero age MS. This mass is in better agreement with the expected value for a B2V star, suggesting that the previous inconsistency in the mass estimate was likely due to its different extinction law, reddening and distance.
        
        \subsubsection{HD 36655 (\#7)}
        
            This star presents apparent emission in the Br13 line, which may lead to misclassification as a B6V star rather than a later type. At a distance of 350\,pc, the estimated mass and age are $M=2.5 \pm 0.5\,M_\odot$ and $4^{+1.0}_{-1.0}$\,Myrs, respectively, which is consistent with the values previously estimated (see Table\,\ref{tab:spec_class}). This suggests that the true spectral type may be later than that inferred from the IR Brackett-line relation, probably a B9V.
        
        \subsubsection{HD 37356 (\#40)}

            This object is a Herbig star characterized by an extinction law of $R_v$ = 3.83, a reddening of E(B-V) = 0.18, and a distance of 450\,pc \citep{Wegner02,Hernandez05}. Adopting these parameters, the estimated mass increases to $M = 10.0 \pm 3.3\,M_\odot$, making it the most massive star in our sample. However, with these revised parameters, this star lies very close to the MS, and its age estimate shows significant discrepancy and large uncertainties. Despite this, we classify it as a peculiar PMS candidate, due to its identification as a Herbig star and its location along the PMS tracks across the CMDs.
        
        \subsubsection{HD 37687 (\#47)}
        
            This star is known to have weak He and strong Si lines \citep{rom21}, which compromise the reliability of optical classification based on the line ratios. Nevertheless, direct comparison indicates a good match with a B5V template, consistent with the IR-based classification. The isochrone-based mass ($M = 4.7 \pm 0.5\,M_\odot$) and age ($0.5^{+0.1}_{-0.1}$\,Myrs) estimates at its distance of 440\,pc are consistent with the previous estimated value and agree with the B5V spectral type mass.
        
        \subsubsection{V1148 Ori (\#53)}
        
            This object is a known magnetic chemically peculiar star with weak He lines \citep{rom21}, classified as a Herbig star. Due to the absence of IR spectra in APOGEE, classification via the Brackett-line relation was not possible. However, the optical spectra indicate, by visual inspection, a B8V type. This is consistent with both the isochrone-based mass estimate ($\sim$ M = 4.2 $\pm$ 1.0\,M$_\odot$) and the photometry, in agreement with the literature \citep{gla19}.
    
\section{Conclusions}
    \label{sec:conclusions}
    
    We estimated the spectral types of B and early A stars surrounding the ON using both IR and optical spectra. In most cases, the two wavelength regimes provide consistent classifications within one subtype, supporting the robustness of our approach. Some of these stars were previously classified as class I--III, but we were able to reclassify them as class V stars, as they exhibit features typical of young stellar objects rather than evolved stars. Furthermore, by means of isochrones fitting, we estimated their masses and ages, finding that several of them lie on the PMS track, reinforcing their youth. This result is consistently seen across the \textit{Gaia}, 2MASS, and WISE CMDs, where the IR planes are particularly informative in the presence of reddening and potential IR excess.
    
    We also derived the typical mass range allowed for each B spectral subtype based on dynamical mass measurements from eclipsing binaries. By fitting Gaussian profiles to the mass distributions per subtype, we obtained mean values and $1\sigma$ dispersions that closely match our isochrone-based masses. This procedure provides an independent constraint on the mass and age estimates from isochrone fitting. The agreement between these independent constraints increases confidence that the adopted isochrones are appropriate for the ON environment.
    
    Using a grid of young isochrones, we identified several PMS candidates in the region. A subset of sources shows spectroscopic variability and/or emission components (notably in Balmer and Brackett lines), suggestive of circumstellar material, magnetic activity, or binarity; these objects are priority targets for time-domain spectroscopy and polarimetry. The IR CMDs were particularly effective for identifying PMS objects, likely due to their reduced sensitivity to extinction and to excess emission from circumstellar environments.

    Based on the derived ages, we observed multiple populations across the age distribution in different evolutionary stages. The formation of younger early-type stars could have been triggered by the feedback from older nearby populations, while the MS and older PMS population may have formed as a consequence of the large-scale events that shaped the Orion complex region \citep{bouy14,grossschedl21}. This distribution provides valuable insight into the star formation scenario in the outskirts of the ON.
    
    In a broader context, our results refine the census of early-type young stars in Orion---a benchmark complex for studying more massive star formation \citep{zinnecker07,krumholz14}. The presence of robust PMS B-star candidates in the ON provides empirical anchors for the short-lived early phases of more massive-star evolution \citep{motte18}. The coexistence of PMS and MS early-type stars in the same field is consistent with a temporally extended or sequential star-formation history in Orion \citep{bouy14,kounkel18,grossschedl21}.
    
    Beyond the ON, the combined strategy that we have employed---optical+IR spectroscopy tied to astrometric/photometric constraints and validated against dynamical masses---offers a viable framework for distinguishing PMS from evolved stars in other star-forming regions. This integrative approach will be crucial for future surveys aimed at mapping the earliest evolutionary pathways of B-type stars and their environments.
        
\section*{acknowledgements}

      The authors acknowledge the Brazilian Agencies FAPEMIG (CEX PPM 00225-09 and PAPG PD\&I nº5.18/2022), CAPES, CNPq (201228/2004-1), and PRPq/UFMG/FUNDEP (14998) for the financial support. Additionally, the authors acknowledge the referee and editor for the valuable comments that helped to improve this work. P. Braz acknowledges the Universidad de La Serena for the support during the technical visit to develop this work. W.J.B. Corradi acknowledges the support from Laborat\'orio Nacional de Astrof\'isica LNA/MCTI. F.Maia acknowledges financial support from CNPq (proc. 404482/2021-0) and FAPERJ (proc. E-26/201.386/2022). E. B. A. acknowledges financial support from the Brazilian National Council for Scientific and Technological Development (CNPq) under Grant No.404160/2025-5 (CNPq/MCTI Call No.44/2024 – Universal) and Universidade Estadual de Feira de Santana for the support received by the Program FINAPESQ (Grant No.051/2025).
    
\section*{Data Availability}

The spectroscopic data used are available in the ESO archive (\text{http://archive.eso.org/eso/eso\_archive\_main.html}) and/or in the APOGEE archive (\text{https://data.sdss.org/sas/dr17/apogee/spectro/redux/dr17/}) and/or in the CASLEO archive (\text{http://179.41.13.34:58880/casleo-db/casleo-db.html}. The photometric data underlying this article is publicly available (\textit{Gaia}, 2MASS and WISE). The data underlying this article are available in the article and in its online supplementary material.
 


\bibliographystyle{mnras}
\bibliography{BPMS} 

@ARTICLE{apogee,
       author = {{Abdurro'uf} and {Accetta}, Katherine and {Aerts}, Conny and {Silva Aguirre}, V{\'\i}ctor and {Ahumada}, Romina and {Ajgaonkar}, Nikhil and {Filiz Ak}, N. and {Alam}, Shadab and {Allende Prieto}, Carlos and {Almeida}, Andr{\'e}s and et al.},
        title = "{The Seventeenth Data Release of the Sloan Digital Sky Surveys: Complete Release of MaNGA, MaStar, and APOGEE-2 Data}",
      journal = {\apjs},
         year = 2022,
        month = apr,
       volume = {259},
       number = {2},
          eid = {35},
        pages = {35},
          doi = {10.3847/1538-4365/ac4414},
archivePrefix = {arXiv},
       eprint = {2112.02026},
 primaryClass = {astro-ph.GA},
       adsurl = {https://ui.adsabs.harvard.edu/abs/2022ApJS..259...35A}
}

@ARTICLE{abt04,
       author = {{Abt}, Helmut A.},
        title = "{Spectral Classification of Stars in A Supplement to the Bright Star Catalogue}",
      journal = {\apjs},
         year = 2004,
        month = nov,
       volume = {155},
       number = {1},
        pages = {175-177},
          doi = {10.1086/423803},
       adsurl = {https://ui.adsabs.harvard.edu/abs/2004ApJS..155..175A}
}

@ARTICLE{AlonsoAlbi09,
       author = {{Alonso-Albi}, T. and {Fuente}, A. and {Bachiller}, R. and {Neri}, R. and {Planesas}, P. and {Testi}, L. and {Bern{\'e}}, O. and {Joblin}, C.},
        title = "{Circumstellar disks around Herbig Be stars}",
      journal = {\aap},
         year = 2009,
        month = apr,
       volume = {497},
       number = {1},
        pages = {117-136},
          doi = {10.1051/0004-6361/200810401},
archivePrefix = {arXiv},
       eprint = {0812.1636},
 primaryClass = {astro-ph},
       adsurl = {https://ui.adsabs.harvard.edu/abs/2009A&A...497..117A}
}

@ARTICLE{alvesbouy12,
       author = {{Alves}, J. and {Bouy}, H.},
        title = "{Orion revisited. I. The massive cluster in front of the Orion nebula cluster}",
      journal = {\aap},
         year = 2012,
        month = nov,
       volume = {547},
          eid = {A97},
        pages = {A97},
          doi = {10.1051/0004-6361/201220119},
archivePrefix = {arXiv},
       eprint = {1209.3787},
 primaryClass = {astro-ph.GA},
       adsurl = {https://ui.adsabs.harvard.edu/abs/2012A&A...547A..97A}
}

@ARTICLE{Amores17,
       author = {{Am{\^o}res}, E.~B. and {Robin}, A.~C. and {Reyl{\'e}}, C.},
        title = "{Evolution over time of the Milky Way's disc shape}",
      journal = {\aap},
         year = 2017,
        month = jun,
       volume = {602},
          eid = {A67},
        pages = {A67},
          doi = {10.1051/0004-6361/201628461},
archivePrefix = {arXiv},
       eprint = {1701.00475},
 primaryClass = {astro-ph.GA},
       adsurl = {https://ui.adsabs.harvard.edu/abs/2017A&A...602A..67A}
}

@INCOLLECTION{bally08,
       author = {{Bally}, J.},
        title = "{Overview of the Orion Complex}",
    booktitle = {Handbook of Star Forming Regions, Volume I},
         year = 2008,
       editor = {{Reipurth}, B.},
       volume = {4},
        pages = {459},
          doi = {10.48550/arXiv.0812.0046},
       adsurl = {https://ui.adsabs.harvard.edu/abs/2008hsf1.book..459B}
}

@ARTICLE{bouy14,
       author = {{Bouy}, H. and {Alves}, J. and {Bertin}, E. and {Sarro}, L.~M. and {Barrado}, D.},
        title = "{Orion revisited. II. The foreground population to Orion A}",
      journal = {\aap},
         year = 2014,
        month = apr,
       volume = {564},
          eid = {A29},
        pages = {A29},
          doi = {10.1051/0004-6361/201323191},
archivePrefix = {arXiv},
       eprint = {1402.1034},
 primaryClass = {astro-ph.SR},
       adsurl = {https://ui.adsabs.harvard.edu/abs/2014A&A...564A..29B}
}

@mastersthesis{braz22,
    author = "{Braz}, P. H. F. B.",
    title = "{Espectroscopia das estrelas jovens HD\,141569, HD\,144432, HD\,163296 e o aglomerado aberto NGC\,1981}",
    school = "{Universidade Federal de Minas Gerais}",
    year = 2022,
    url = "http://hdl.handle.net/1843/61547"
}

@ARTICLE{bre12,
       author = {{Bressan}, Alessandro and {Marigo}, Paola and {Girardi}, L{\'e}o. and {Salasnich}, Bernardo and {Dal Cero}, Claudia and {Rubele}, Stefano and {Nanni}, Ambra},
        title = "{PARSEC: stellar tracks and isochrones with the PAdova and TRieste Stellar Evolution Code}",
      journal = {\mnras},
         year = 2012,
        month = nov,
       volume = {427},
       number = {1},
        pages = {127-145},
          doi = {10.1111/j.1365-2966.2012.21948.x},
archivePrefix = {arXiv},
       eprint = {1208.4498},
 primaryClass = {astro-ph.SR},
       adsurl = {https://ui.adsabs.harvard.edu/abs/2012MNRAS.427..127B}
}

@ARTICLE{Brittain23,
       author = {{Brittain}, Sean D. and {Kamp}, Inga and {Meeus}, Gwendolyn and {Oudmaijer}, Ren{\'e} D. and {Waters}, L.~B.~F.~M.},
        title = "{Herbig Stars}",
      journal = {\ssr},
         year = 2023,
        month = feb,
       volume = {219},
       number = {1},
          eid = {7},
        pages = {7},
          doi = {10.1007/s11214-023-00949-z},
archivePrefix = {arXiv},
       eprint = {2301.01165},
 primaryClass = {astro-ph.SR},
       adsurl = {https://ui.adsabs.harvard.edu/abs/2023SSRv..219....7B}
}

@ARTICLE{cab19,
       author = {{Caballero}, J.~A. and {de Burgos}, A. and {Alonso-Floriano}, F.~J. and {Cabrera-Lavers}, A. and {Garc{\'\i}a-{\'A}lvarez}, D. and {Montes}, D.},
        title = "{Stars and brown dwarfs in the {\ensuremath{\sigma}} Orionis cluster. IV. IDS/INT and OSIRIS/GTC spectroscopy and Gaia DR2 astrometry}",
      journal = {\aap},
         year = 2019,
        month = sep,
       volume = {629},
          eid = {A114},
        pages = {A114},
          doi = {10.1051/0004-6361/201935987},
archivePrefix = {arXiv},
       eprint = {1908.10340},
 primaryClass = {astro-ph.SR},
       adsurl = {https://ui.adsabs.harvard.edu/abs/2019A&A...629A.114C}
}

@ARTICLE{Campbell23,
       author = {{Campbell}, Hunter and {Khilfeh}, Elliott and {Covey}, Kevin R. and {Kounkel}, Marina and {Ballantyne}, Richard and {Corey}, Sabrina and {Rom{\'a}n-Z{\'u}{\~n}iga}, Carlos G. and {Hern{\'a}ndez}, Jes{\'u}s and {Manzo Mart{\'\i}nez}, Ezequiel and {Pe{\~n}a Ram{\'\i}rez}, Karla and et al.},
        title = "{Pre-main-sequence Brackett Emitters in the APOGEE DR17 Catalog: Line Strengths and Physical Properties of Accretion Columns}",
      journal = {\apj},
         year = 2023,
        month = jan,
       volume = {942},
       number = {1},
          eid = {22},
        pages = {22},
          doi = {10.3847/1538-4357/aca324},
archivePrefix = {arXiv},
       eprint = {2211.06454},
 primaryClass = {astro-ph.SR},
       adsurl = {https://ui.adsabs.harvard.edu/abs/2023ApJ...942...22C}
}

@ARTICLE{Cardelli89,
       author = {{Cardelli}, Jason A. and {Clayton}, Geoffrey C. and {Mathis}, John S.},
        title = "{The Relationship between Infrared, Optical, and Ultraviolet Extinction}",
      journal = {\apj},
         year = 1989,
        month = oct,
       volume = {345},
        pages = {245},
          doi = {10.1086/167900},
       adsurl = {https://ui.adsabs.harvard.edu/abs/1989ApJ...345..245C}
}

@ARTICLE{cho15,
       author = {{Chojnowski}, S. Drew and {Whelan}, David G. and {Wisniewski}, John P. and {Majewski}, Steven R. and {Hall}, Matthew and {Shetrone}, Matthew and {Beaton}, Rachael and {Burton}, Adam and {Damke}, Guillermo and {Eikenberry}, Steve and et al.},
        title = "{High-Resolution H-Band Spectroscopy of Be Stars With SDSS-III/Apogee: I. New Be Stars, Line Identifications, and Line Profiles}",
      journal = {\aj},
         year = 2015,
        month = jan,
       volume = {149},
       number = {1},
          eid = {7},
        pages = {7},
          doi = {10.1088/0004-6256/149/1/7},
archivePrefix = {arXiv},
       eprint = {1409.4668},
 primaryClass = {astro-ph.SR},
       adsurl = {https://ui.adsabs.harvard.edu/abs/2015AJ....149....7C}
}

@ARTICLE{Derkink24,
       author = {{Derkink}, A.~R. and {Ram{\'\i}rez-Tannus}, M.~C. and {Kaper}, L. and {de Koter}, A. and {Backs}, F. and {Poorta}, J. and {van Gelder}, M.~L.},
        title = "{Spectroscopic variability of massive pre-main-sequence stars in M17}",
      journal = {\aap},
         year = 2024,
        month = jan,
       volume = {681},
          eid = {A112},
        pages = {A112},
          doi = {10.1051/0004-6361/202347369},
archivePrefix = {arXiv},
       eprint = {2310.04287},
 primaryClass = {astro-ph.SR},
       adsurl = {https://ui.adsabs.harvard.edu/abs/2024A&A...681A.112D}
}

@ARTICLE{Dias21,
       author = {{Dias}, W.~S. and {Monteiro}, H. and {Moitinho}, A. and {L{\'e}pine}, J.~R.~D. and {Carraro}, G. and {Paunzen}, E. and {Alessi}, B. and {Villela}, L.},
        title = "{Updated parameters of 1743 open clusters based on Gaia DR2}",
      journal = {\mnras},
         year = 2021,
        month = jun,
       volume = {504},
       number = {1},
        pages = {356-371},
          doi = {10.1093/mnras/stab770},
archivePrefix = {arXiv},
       eprint = {2103.12829},
 primaryClass = {astro-ph.SR},
       adsurl = {https://ui.adsabs.harvard.edu/abs/2021MNRAS.504..356D}
}

@ARTICLE{Edenhofer24,
       author = {{Edenhofer}, Gordian and {Zucker}, Catherine and {Frank}, Philipp and {Saydjari}, Andrew K. and {Speagle}, Joshua S. and {Finkbeiner}, Douglas and {En{\ss}lin}, Torsten A.},
        title = "{A parsec-scale Galactic 3D dust map out to 1.25 kpc from the Sun}",
      journal = {\aap},
         year = 2024,
        month = may,
       volume = {685},
          eid = {A82},
        pages = {A82},
          doi = {10.1051/0004-6361/202347628},
archivePrefix = {arXiv},
       eprint = {2308.01295},
 primaryClass = {astro-ph.GA},
       adsurl = {https://ui.adsabs.harvard.edu/abs/2024A&A...685A..82E}
}

@ARTICLE{eker18,
       author = {{Eker}, Z. and {Bak{\i}{\c{s}}}, V. and {Bilir}, S. and {Soydugan}, F. and {Steer}, I. and {Soydugan}, E. and {Bak{\i}{\c{s}}}, H. and {Ali{\c{c}}avu{\c{s}}}, F. and {Aslan}, G. and {Alpsoy}, M.},
        title = "{Interrelated main-sequence mass-luminosity, mass-radius, and mass-effective temperature relations}",
      journal = {\mnras},
         year = 2018,
        month = oct,
       volume = {479},
       number = {4},
        pages = {5491-5511},
          doi = {10.1093/mnras/sty1834},
archivePrefix = {arXiv},
       eprint = {1807.02568},
 primaryClass = {astro-ph.SR},
       adsurl = {https://ui.adsabs.harvard.edu/abs/2018MNRAS.479.5491E}
}

@ARTICLE{FinkenzellerMundt84,
       author = {{Finkenzeller}, U. and {Mundt}, R.},
        title = "{The Herbig Ae/Be stars associated with nebulosity.}",
      journal = {\aaps},
         year = 1984,
        month = jan,
       volume = {55},
        pages = {109-141},
       adsurl = {https://ui.adsabs.harvard.edu/abs/1984A&AS...55..109F}
}

@ARTICLE{gaiaDR3,
       author = {{Gaia Collaboration} and {Vallenari}, A. and {Brown}, A.~G.~A. and {Prusti}, T. and {de Bruijne}, J.~H.~J. and {Arenou}, F. and {Babusiaux}, C. and {Biermann}, M. and {Creevey}, O.~L. and {Ducourant}, C. and et al.},
        title = "{Gaia Data Release 3. Summary of the content and survey properties}",
      journal = {\aap},
         year = 2023,
        month = jun,
       volume = {674},
          eid = {A1},
        pages = {A1},
          doi = {10.1051/0004-6361/202243940},
archivePrefix = {arXiv},
       eprint = {2208.00211},
 primaryClass = {astro-ph.GA},
       adsurl = {https://ui.adsabs.harvard.edu/abs/2023A&A...674A...1G}
}

@ARTICLE{gla19,
       author = {{Glagolevskij}, Yu. V.},
        title = "{On Properties of Main Sequence Magnetic Stars}",
      journal = {Astrophysical Bulletin},
         year = 2019,
        month = jan,
       volume = {74},
       number = {1},
        pages = {66-79},
          doi = {10.1134/S1990341319010073},
       adsurl = {https://ui.adsabs.harvard.edu/abs/2019AstBu..74...66G}
}

@BOOK{graycorbally09,
       author = {{Gray}, Richard O. and {Corbally}, J., Christopher},
        title = "{Stellar Spectral Classification}",
         year = 2009,
       adsurl = {https://ui.adsabs.harvard.edu/abs/2009ssc..book.....G}
}

@ARTICLE{Grossschedl19,
       author = {{Gro{\ss}schedl}, Josefa Elisabeth and {Alves}, Jo{\~a}o and {Teixeira}, Paula S. and {Bouy}, Herv{\'e} and {Forbrich}, Jan and {Lada}, Charles J. and {Meingast}, Stefan and {Hacar}, {\'A}lvaro and {Ascenso}, Joana and {Ackerl}, Christine and et al.},
        title = "{VISION - Vienna survey in Orion. III. Young stellar objects in Orion A}",
      journal = {\aap},
         year = 2019,
        month = feb,
       volume = {622},
          eid = {A149},
        pages = {A149},
          doi = {10.1051/0004-6361/201832577},
archivePrefix = {arXiv},
       eprint = {1810.00878},
 primaryClass = {astro-ph.SR},
       adsurl = {https://ui.adsabs.harvard.edu/abs/2019A&A...622A.149G}
}

@ARTICLE{grossschedl21,
       author = {{Gro{\ss}schedl}, Josefa E. and {Alves}, Jo{\~a}o and {Meingast}, Stefan and {Herbst-Kiss}, Gabor},
        title = "{3D dynamics of the Orion cloud complex. Discovery of coherent radial gas motions at the 100-pc scale}",
      journal = {\aap},
         year = 2021,
        month = mar,
       volume = {647},
          eid = {A91},
        pages = {A91},
          doi = {10.1051/0004-6361/202038913},
archivePrefix = {arXiv},
       eprint = {2007.07254},
 primaryClass = {astro-ph.SR},
       adsurl = {https://ui.adsabs.harvard.edu/abs/2021A&A...647A..91G}
}

@ARTICLE{Guimaraes06,
       author = {{Guimar{\~a}es}, M.~M. and {Alencar}, S.~H.~P. and {Corradi}, W.~J.~B. and {Vieira}, S.~L.~A.},
        title = "{Stellar parameters and evidence of circumstellar activity for a sample of Herbig Ae/Be stars}",
      journal = {\aap},
         year = 2006,
        month = oct,
       volume = {457},
       number = {2},
        pages = {581-589},
          doi = {10.1051/0004-6361:20065005},
       adsurl = {https://ui.adsabs.harvard.edu/abs/2006A&A...457..581G}
}

@ARTICLE{herbig60,
       author = {{Herbig}, George H.},
        title = "{The Spectra of Be- and Ae-Type Stars Associated with Nebulosity}",
      journal = {\apjs},
         year = 1960,
        month = mar,
       volume = {4},
        pages = {337},
          doi = {10.1086/190050},
       adsurl = {https://ui.adsabs.harvard.edu/abs/1960ApJS....4..337H}
}

@ARTICLE{Hernandez05,
       author = {{Hern{\'a}ndez}, Jes{\'u}s and {Calvet}, Nuria and {Hartmann}, Lee and {Brice{\~n}o}, C{\'e}sar and {Sicilia-Aguilar}, Aurora and {Berlind}, Perry},
        title = "{Herbig Ae/Be Stars in nearby OB Associations}",
      journal = {\aj},
         year = 2005,
        month = feb,
       volume = {129},
       number = {2},
        pages = {856-871},
          doi = {10.1086/426918},
archivePrefix = {arXiv},
       eprint = {astro-ph/0410494},
 primaryClass = {astro-ph},
       adsurl = {https://ui.adsabs.harvard.edu/abs/2005AJ....129..856H}
}

@BOOK{houk99,
       author = {{Houk}, Nancy and {Swift}, Carrie},
        title = "{Michigan catalogue of two-dimensional spectral types for the HD Stars ; vol. 5}",
         year = 1999,
       volume = {5},
       publisher = {Department of Astronomy, University of Michigan},
        series    = {Michigan Catalogue of Two-dimensional Spectral Types},
        address   = {Ann Arbor},
       adsurl = {https://ui.adsabs.harvard.edu/abs/1999mctd.book.....H}
}

@ARTICLE{hsu13,
       author = {{Hsu}, Wen-Hsin and {Hartmann}, Lee and {Allen}, Lori and {Hern{\'a}ndez}, Jes{\'u}s and {Megeath}, S.~T. and {Tobin}, John J. and {Ingleby}, Laura},
        title = "{Evidence for Environmental Dependence of the Upper Stellar Initial Mass Function in Orion A}",
      journal = {\apj},
         year = 2013,
        month = feb,
       volume = {764},
       number = {2},
          eid = {114},
        pages = {114},
          doi = {10.1088/0004-637X/764/2/114},
archivePrefix = {arXiv},
       eprint = {1212.1171},
 primaryClass = {astro-ph.SR},
       adsurl = {https://ui.adsabs.harvard.edu/abs/2013ApJ...764..114H}
}

@ARTICLE{kue24,
       author = {{Kue{\ss}}, L. and {Paunzen}, E. and {Faltov{\'a}}, N. and {Jadlovsk{\'y}}, D. and {Labaj}, M. and {Mesar{\v{c}}}, M. and {Mondal}, P. and {Pri{\v{s}}egen}, M. and {Ramezani}, T. and {Sup{\'\i}kov{\'a}}, J. and et al.},
        title = "{Chemically peculiar stars on the pre-main sequence}",
      journal = {\aap},
         year = 2024,
        month = jul,
       volume = {687},
          eid = {A176},
        pages = {A176},
          doi = {10.1051/0004-6361/202348926},
archivePrefix = {arXiv},
       eprint = {2405.08946},
 primaryClass = {astro-ph.SR},
       adsurl = {https://ui.adsabs.harvard.edu/abs/2024A&A...687A.176K}
}

@ARTICLE{ker19,
       author = {{Kervella}, Pierre and {Arenou}, Fr{\'e}d{\'e}ric and {Mignard}, Fran{\c{c}}ois and {Th{\'e}venin}, Fr{\'e}d{\'e}ric},
        title = "{Stellar and substellar companions of nearby stars from Gaia DR2. Binarity from proper motion anomaly}",
      journal = {\aap},
         year = 2019,
        month = mar,
       volume = {623},
          eid = {A72},
        pages = {A72},
          doi = {10.1051/0004-6361/201834371},
archivePrefix = {arXiv},
       eprint = {1811.08902},
 primaryClass = {astro-ph.SR},
       adsurl = {https://ui.adsabs.harvard.edu/abs/2019A&A...623A..72K}
}

@ARTICLE{kounkel18,
       author = {{Kounkel}, Marina and {Covey}, Kevin and {Su{\'a}rez}, Genaro and {Rom{\'a}n-Z{\'u}{\~n}iga}, Carlos and {Hernandez}, Jesus and {Stassun}, Keivan and {Jaehnig}, Karl O. and {Feigelson}, Eric D. and {Pe{\~n}a Ram{\'\i}rez}, Karla and {Roman-Lopes}, Alexandre and et al.},
        title = "{The APOGEE-2 Survey of the Orion Star-forming Complex. II. Six-dimensional Structure}",
      journal = {\aj},
         year = 2018,
        month = sep,
       volume = {156},
       number = {3},
          eid = {84},
        pages = {84},
          doi = {10.3847/1538-3881/aad1f1},
archivePrefix = {arXiv},
       eprint = {1805.04649},
 primaryClass = {astro-ph.SR},
       adsurl = {https://ui.adsabs.harvard.edu/abs/2018AJ....156...84K}
}

@ARTICLE{krumholz14,
       author = {{Krumholz}, Mark R.},
        title = "{The big problems in star formation: The star formation rate, stellar clustering, and the initial mass function}",
      journal = {\physrep},
         year = 2014,
        month = jun,
       volume = {539},
        pages = {49-134},
          doi = {10.1016/j.physrep.2014.02.001},
archivePrefix = {arXiv},
       eprint = {1402.0867},
 primaryClass = {astro-ph.GA},
       adsurl = {https://ui.adsabs.harvard.edu/abs/2014PhR...539...49K}
}

@ARTICLE{Kuhn19,
       author = {{Kuhn}, Michael A. and {Hillenbrand}, Lynne A. and {Sills}, Alison and {Feigelson}, Eric D. and {Getman}, Konstantin V.},
        title = "{Kinematics in Young Star Clusters and Associations with Gaia DR2}",
      journal = {\apj},
         year = 2019,
        month = jan,
       volume = {870},
       number = {1},
          eid = {32},
        pages = {32},
          doi = {10.3847/1538-4357/aaef8c},
archivePrefix = {arXiv},
       eprint = {1807.02115},
 primaryClass = {astro-ph.GA},
       adsurl = {https://ui.adsabs.harvard.edu/abs/2019ApJ...870...32K}
}

@ARTICLE{kyr22,
       author = {{Kyritsis}, E. and {Maravelias}, G. and {Zezas}, A. and {Bonfini}, P. and {Kovlakas}, K. and {Reig}, P.},
        title = "{A new automated tool for the spectral classification of OB stars}",
      journal = {\aap},
         year = 2022,
        month = jan,
       volume = {657},
          eid = {A62},
        pages = {A62},
          doi = {10.1051/0004-6361/202040224},
archivePrefix = {arXiv},
       eprint = {2110.10669},
 primaryClass = {astro-ph.SR},
       adsurl = {https://ui.adsabs.harvard.edu/abs/2022A&A...657A..62K}
}

@ARTICLE{LAMOST,
       author = {{Cui}, Xiang-Qun and {Zhao}, Yong-Heng and {Chu}, Yao-Quan and {Li}, Guo-Ping and {Li}, Qi and {Zhang}, Li-Ping and {Su}, Hong-Jun and {Yao}, Zheng-Qiu and {Wang}, Ya-Nan and {Xing}, Xiao-Zheng and et al.},
        title = "{The Large Sky Area Multi-Object Fiber Spectroscopic Telescope (LAMOST)}",
      journal = {Research in Astronomy and Astrophysics},
         year = 2012,
        month = sep,
       volume = {12},
       number = {9},
        pages = {1197-1242},
          doi = {10.1088/1674-4527/12/9/003},
       adsurl = {https://ui.adsabs.harvard.edu/abs/2012RAA....12.1197C}
}

@ARTICLE{maia10,
       author = {{Maia}, F.~F.~S. and {Corradi}, W.~J.~B. and {Santos}, Jr., J.~F.~C.},
        title = "{Characterization and photometric membership of the open cluster NGC1981}",
      journal = {\mnras},
         year = 2010,
        month = sep,
       volume = {407},
       number = {3},
        pages = {1875-1886},
          doi = {10.1111/j.1365-2966.2010.17034.x},
archivePrefix = {arXiv},
       eprint = {1005.3047},
 primaryClass = {astro-ph.SR},
       adsurl = {https://ui.adsabs.harvard.edu/abs/2010MNRAS.407.1875M}
}

@ARTICLE{mar23,
       author = {{Marton}, G{\'a}bor and {{\'A}brah{\'a}m}, P{\'e}ter and {Rimoldini}, Lorenzo and {Audard}, Marc and {Kun}, M{\'a}ria and {Nagy}, Zs{\'o}fia and {K{\'o}sp{\'a}l}, {\'A}gnes and {Szabados}, L{\'a}szl{\'o} and {Holl}, Berry and {Gavras}, Panagiotis and et al.},
        title = "{Gaia Data Release 3. Validating the classification of variable young stellar object candidates}",
      journal = {\aap},
         year = 2023,
        month = jun,
       volume = {674},
          eid = {A21},
        pages = {A21},
          doi = {10.1051/0004-6361/202244101},
archivePrefix = {arXiv},
       eprint = {2206.05796},
 primaryClass = {astro-ph.SR},
       adsurl = {https://ui.adsabs.harvard.edu/abs/2023A&A...674A..21M}
}

@ARTICLE{megeath16,
       author = {{Megeath}, S.~T. and {Gutermuth}, R. and {Muzerolle}, J. and {Kryukova}, E. and {Hora}, J.~L. and {Allen}, L.~E. and {Flaherty}, K. and {Hartmann}, L. and {Myers}, P.~C. and {Pipher}, J.~L. and et al.},
        title = "{The Spitzer Space Telescope Survey of the Orion A and B Molecular Clouds. II. The Spatial Distribution and Demographics of Dusty Young Stellar Objects}",
      journal = {\aj},
         year = 2016,
        month = jan,
       volume = {151},
       number = {1},
          eid = {5},
        pages = {5},
          doi = {10.3847/0004-6256/151/1/5},
archivePrefix = {arXiv},
       eprint = {1511.01202},
 primaryClass = {astro-ph.GA},
       adsurl = {https://ui.adsabs.harvard.edu/abs/2016AJ....151....5M}
}

@ARTICLE{morgan66,
       author = {{Morgan}, W.~W. and {Lod{\'e}n}, Kerstin},
        title = "{Some characteristics of the orion association}",
      journal = {Vistas in Astronomy},
         year = 1966,
        month = jan,
       volume = {8},
       number = {1},
        pages = {83-88},
          doi = {10.1016/0083-6656(66)90023-7},
       adsurl = {https://ui.adsabs.harvard.edu/abs/1966VA......8...83M}
}

@ARTICLE{motte18,
       author = {{Motte}, Fr{\'e}d{\'e}rique and {Bontemps}, Sylvain and {Louvet}, Fabien},
        title = "{High-Mass Star and Massive Cluster Formation in the Milky Way}",
      journal = {\araa},
         year = 2018,
        month = sep,
       volume = {56},
        pages = {41-82},
          doi = {10.1146/annurev-astro-091916-055235},
archivePrefix = {arXiv},
       eprint = {1706.00118},
 primaryClass = {astro-ph.GA},
       adsurl = {https://ui.adsabs.harvard.edu/abs/2018ARA&A..56...41M}
}

@ARTICLE{murphy20,
       author = {{Murphy}, Simon J. and {Gray}, Richard O. and {Corbally}, Christopher J. and {Kuehn}, Charles and {Bedding}, Timothy R. and {Killam}, Josiah},
        title = "{The discovery of lambda Bootis stars - the Southern Survey II}",
      journal = {\mnras},
         year = 2020,
        month = dec,
       volume = {499},
       number = {2},
        pages = {2701-2713},
          doi = {10.1093/mnras/staa2347},
archivePrefix = {arXiv},
       eprint = {2008.02392},
 primaryClass = {astro-ph.SR},
       adsurl = {https://ui.adsabs.harvard.edu/abs/2020MNRAS.499.2701M}
}

@INPROCEEDINGS{Natta00,
       author = {{Natta}, A. and {Grinin}, V. and {Mannings}, V.},
        title = "{Properties and Evolution of Disks around Pre-Main-Sequence Stars of Intermediate Mass}",
    booktitle = {Protostars and Planets IV},
         year = 2000,
       editor = {{Mannings}, V. and {Boss}, A.~P. and {Russell}, S.~S.},
        month = may,
        pages = {559-588},
       adsurl = {https://ui.adsabs.harvard.edu/abs/2000prpl.conf..559N}
}

@ARTICLE{ner24,
       author = {{Negueruela}, I. and {Sim{\'o}n-D{\'\i}az}, S. and {de Burgos}, A. and {Casasbuenas}, A. and {Beck}, P.~G.},
        title = "{The IACOB project: XII. New grid of northern standards for the spectral classification of B-type stars}",
      journal = {\aap},
         year = 2024,
        month = oct,
       volume = {690},
          eid = {A176},
        pages = {A176},
          doi = {10.1051/0004-6361/202449298},
archivePrefix = {arXiv},
       eprint = {2407.04163},
 primaryClass = {astro-ph.SR},
       adsurl = {https://ui.adsabs.harvard.edu/abs/2024A&A...690A.176N}
}

@ARTICLE{Ochsendorf11,
       author = {{Ochsendorf}, B.~B. and {Ellerbroek}, L.~E. and {Chini}, R. and {Hartoog}, O.~E. and {Hoffmeister}, V. and {Waters}, L.~B.~F.~M. and {Kaper}, L.},
        title = "{First firm spectral classification of an early-B pre-main-sequence star: B275 in <ASTROBJ>M 17</ASTROBJ>}",
      journal = {\aap},
         year = 2011,
        month = dec,
       volume = {536},
          eid = {L1},
        pages = {L1},
          doi = {10.1051/0004-6361/201118089},
archivePrefix = {arXiv},
       eprint = {1110.5755},
 primaryClass = {astro-ph.SR},
       adsurl = {https://ui.adsabs.harvard.edu/abs/2011A&A...536L...1O}
}

@ARTICLE{Pickles98,
       author = {{Pickles}, A.~J.},
        title = "{A Stellar Spectral Flux Library: 1150-25000 {\r{A}}}",
      journal = {\pasp},
         year = 1998,
        month = jul,
       volume = {110},
       number = {749},
        pages = {863-878},
          doi = {10.1086/316197},
       adsurl = {https://ui.adsabs.harvard.edu/abs/1998PASP..110..863P}
}

@ARTICLE{dragons,
       author = {{Placco}, Vinicius M. and {Herrera}, David and {Merino}, Brian M. and {US National Gemini Office} and {Hirst}, Paul and {Labrie}, Kathleen and {Simpson}, Chris and {Turner}, James and {Vacca}, William D. and {Gemini Science User Support Department} and et al.},
        title = "{GHOST Reduced Data Products for the Gemini Observatory Community and Beyond}",
      journal = {Research Notes of the American Astronomical Society},
         year = 2024,
        month = dec,
       volume = {8},
       number = {12},
          eid = {312},
        pages = {312},
          doi = {10.3847/2515-5172/ad9f5f},
archivePrefix = {arXiv},
       eprint = {2412.13239},
 primaryClass = {astro-ph.IM},
       adsurl = {https://ui.adsabs.harvard.edu/abs/2024RNAAS...8..312P}
}

@ARTICLE{Pogodin21,
       author = {{Pogodin}, Mikhail and {Drake}, Natalia and {Beskrovnaya}, Nina and {Pavlovskiy}, Sergei and {Hubrig}, Swetlana and {Sch{\"o}ller}, Markus and {J{\"a}rvinen}, Silva and {Kozlova}, Olesya and {Alekseev}, Ilya},
        title = "{Searching for Magnetospheres around Herbig Ae/Be Stars}",
      journal = {Universe},
         year = 2021,
        month = dec,
       volume = {7},
       number = {12},
        pages = {489},
          doi = {10.3390/universe7120489},
       adsurl = {https://ui.adsabs.harvard.edu/abs/2021Univ....7..489P}
}

@ARTICLE{Quintana25,
       author = {{Quintana}, Alexis L. and {Wright}, Nicholas J. and {Mart{\'\i}nez Garc{\'\i}a}, Juan},
        title = "{A census of OB stars within 1 kpc and the star formation and core collapse supernova rates of the Milky Way}",
      journal = {\mnras},
         year = 2025,
        month = apr,
       volume = {538},
       number = {3},
        pages = {1367-1383},
          doi = {10.1093/mnras/staf083},
archivePrefix = {arXiv},
       eprint = {2503.08286},
 primaryClass = {astro-ph.SR},
       adsurl = {https://ui.adsabs.harvard.edu/abs/2025MNRAS.538.1367Q}
}

@ARTICLE{ramirez20,
       author = {{Ram{\'\i}rez-Preciado}, Valeria G. and {Roman-Lopes}, Alexandre and {Rom{\'a}n-Z{\'u}{\~n}iga}, Carlos G. and {Hern{\'a}ndez}, Jes{\'u}s and {Garc{\'\i}a-Hern{\'a}ndez}, D.~A. and {Stassun}, Keivan and {Stringfellow}, Guy S. and {Kim}, Jinyoung Serena},
        title = "{Spectral Classification of B Stars: The Empirical Sequence Using SDSS-IV/APOGEE Near-IR Data}",
      journal = {\apj},
         year = 2020,
        month = may,
       volume = {894},
       number = {1},
          eid = {5},
        pages = {5},
          doi = {10.3847/1538-4357/ab8127},
archivePrefix = {arXiv},
       eprint = {2003.09469},
 primaryClass = {astro-ph.SR},
       adsurl = {https://ui.adsabs.harvard.edu/abs/2020ApJ...894....5R}
}

@ARTICLE{ramirezTannus17,
       author = {{Ram{\'\i}rez-Tannus}, M.~C. and {Kaper}, L. and {de Koter}, A. and {Tramper}, F. and {Bik}, A. and {Ellerbroek}, L.~E. and {Ochsendorf}, B.~B. and {Ram{\'\i}rez-Agudelo}, O.~H. and {Sana}, H.},
        title = "{Massive pre-main-sequence stars in M17}",
      journal = {\aap},
         year = 2017,
        month = aug,
       volume = {604},
          eid = {A78},
        pages = {A78},
          doi = {10.1051/0004-6361/201629503},
archivePrefix = {arXiv},
       eprint = {1704.08216},
 primaryClass = {astro-ph.SR},
       adsurl = {https://ui.adsabs.harvard.edu/abs/2017A&A...604A..78R}
}

@ARTICLE{rom21,
       author = {{Romanyuk}, I.~I. and {Semenko}, E.~A. and {Moiseeva}, A.~V. and {Yakunin}, I.~A. and {Kudryavtsev}, D.~O.},
        title = "{Magnetic Fields of CP Stars in the Orion OB1 Association. IV. Stars of Subgroup 1b}",
      journal = {Astrophysical Bulletin},
         year = 2021,
        month = jan,
       volume = {76},
       number = {1},
        pages = {39-54},
          doi = {10.1134/S1990341321010090},
archivePrefix = {arXiv},
       eprint = {2104.02434},
 primaryClass = {astro-ph.SR},
       adsurl = {https://ui.adsabs.harvard.edu/abs/2021AstBu..76...39R}
}

@ARTICLE{rum23,
       author = {{Rustem}, Abdurepqet and {L{\"u}}, Guo-Liang and {Liu}, Jin-Zhong and {Zhu}, Chun-Hua and {Zhang}, Yu and {Shen}, Dong-Xiang and {Zhang}, Yu-Hao and {He}, Xiao-Long},
        title = "{A Catalog and Statistical Analysis for Magnetic Stars}",
      journal = {Research in Astronomy and Astrophysics},
         year = 2023,
        month = sep,
       volume = {23},
       number = {9},
          eid = {095024},
        pages = {095024},
          doi = {10.1088/1674-4527/ace9b0},
archivePrefix = {arXiv},
       eprint = {2307.12315},
 primaryClass = {astro-ph.SR},
       adsurl = {https://ui.adsabs.harvard.edu/abs/2023RAA....23i5024R}
}

@ARTICLE{SanchezSanjuan24,
       author = {{S{\'a}nchez-Sanju{\'a}n}, Sergio and {Hern{\'a}ndez}, Jes{\'u}s and {P{\'e}rez-Villegas}, {\'A}ngeles and {Rom{\'a}n-Z{\'u}{\~n}iga}, Carlos and {Aguilar}, Luis and {Ballesteros-Paredes}, Javier and {Bonilla-Barroso}, Andrea},
        title = "{Kinematic study of the Orion Complex: analysing the young stellar clusters from big and small structures}",
      journal = {\mnras},
         year = 2024,
        month = nov,
       volume = {534},
       number = {3},
        pages = {2566-2584},
          doi = {10.1093/mnras/stae2157},
archivePrefix = {arXiv},
       eprint = {2409.09206},
 primaryClass = {astro-ph.GA},
       adsurl = {https://ui.adsabs.harvard.edu/abs/2024MNRAS.534.2566S}
}

@misc{Skiff09,
       author = {{Skiff}, B.~A.},
        title = "{VizieR Online Data Catalog: Catalogue of stellar spectral classifications.}",
 howpublished = {CDS/ADC Collection of Electronic Catalogues, 1, 2023 (2009)},
         year = 2009,
        month = jan,
       adsurl = {https://ui.adsabs.harvard.edu/abs/2009yCat....1.2023S}
}

@ARTICLE{2mass,
       author = {{Skrutskie}, M.~F. and {Cutri}, R.~M. and {Stiening}, R. and {Weinberg}, M.~D. and {Schneider}, S. and {Carpenter}, J.~M. and {Beichman}, C. and {Capps}, R. and {Chester}, T. and {Elias}, J. and et al.},
        title = "{The Two Micron All Sky Survey (2MASS)}",
      journal = {\aj},
         year = 2006,
        month = feb,
       volume = {131},
       number = {2},
        pages = {1163-1183},
          doi = {10.1086/498708},
       adsurl = {https://ui.adsabs.harvard.edu/abs/2006AJ....131.1163S}
}

@ARTICLE{Sprague22,
       author = {{Sprague}, Dani and {Culhane}, Connor and {Kounkel}, Marina and {Olney}, Richard and {Covey}, K.~R. and {Hutchinson}, Brian and {Lingg}, Ryan and {Stassun}, Keivan G. and {Rom{\'a}n-Z{\'u}{\~n}iga}, Carlos G. and {Roman-Lopes}, Alexandre and {Nidever}, David and {Beaton}, Rachael L. and {Borissova}, Jura and {Stutz}, Amelia and {Stringfellow}, Guy S. and {Ram{\'\i}rez}, Karla Pe{\~n}a and {Ram{\'\i}rez-Preciado}, Valeria and {Hern{\'a}ndez}, Jes{\'u}s and {Kim}, Jinyoung Serena and {Lane}, Richard R.},
        title = "{APOGEE Net: An Expanded Spectral Model of Both Low-mass and High-mass Stars}",
      journal = {\aj},
         year = 2022,
        month = apr,
       volume = {163},
       number = {4},
          eid = {152},
        pages = {152},
          doi = {10.3847/1538-3881/ac4de7},
archivePrefix = {arXiv},
       eprint = {2201.03661},
 primaryClass = {astro-ph.GA},
       adsurl = {https://ui.adsabs.harvard.edu/abs/2022AJ....163..152S}
}

@ARTICLE{sota11,
       author = {{Sota}, A. and {Ma{\'\i}z Apell{\'a}niz}, J. and {Walborn}, N.~R. and {Alfaro}, E.~J. and {Barb{\'a}}, R.~H. and {Morrell}, N.~I. and {Gamen}, R.~C. and {Arias}, J.~I.},
        title = "{The Galactic O-Star Spectroscopic Survey. I. Classification System and Bright Northern Stars in the Blue-violet at R \raisebox{-0.5ex}\textasciitilde 2500}",
      journal = {\apjs},
         year = 2011,
        month = apr,
       volume = {193},
       number = {2},
          eid = {24},
        pages = {24},
          doi = {10.1088/0067-0049/193/2/24},
archivePrefix = {arXiv},
       eprint = {1101.4002},
 primaryClass = {astro-ph.GA},
       adsurl = {https://ui.adsabs.harvard.edu/abs/2011ApJS..193...24S}
}

@ARTICLE{The94,
       author = {{The}, P.~S. and {de Winter}, D. and {Perez}, M.~R.},
        title = "{A new catalogue of members and candidate members of the Herbig Ae/Be (HAEBE) stellar group}",
      journal = {\aaps},
         year = 1994,
        month = apr,
       volume = {104},
        pages = {315-339},
       adsurl = {https://ui.adsabs.harvard.edu/abs/1994A&AS..104..315T}
}

@ARTICLE{Vergely22,
       author = {{Vergely}, J.~L. and {Lallement}, R. and {Cox}, N.~L.~J.},
        title = "{Three-dimensional extinction maps: Inverting inter-calibrated extinction catalogues}",
      journal = {\aap},
         year = 2022,
        month = aug,
       volume = {664},
          eid = {A174},
        pages = {A174},
          doi = {10.1051/0004-6361/202243319},
archivePrefix = {arXiv},
       eprint = {2205.09087},
 primaryClass = {astro-ph.GA},
       adsurl = {https://ui.adsabs.harvard.edu/abs/2022A&A...664A.174V}
}

@ARTICLE{vieira03,
       author = {{Vieira}, S.~L.~A. and {Corradi}, W.~J.~B. and {Alencar}, S.~H.~P. and {Mendes}, L.~T.~S. and {Torres}, C.~A.~O. and {Quast}, G.~R. and {Guimar{\~a}es}, M.~M. and {da Silva}, L.},
        title = "{Investigation of 131 Herbig Ae/Be Candidate Stars}",
      journal = {\aj},
         year = 2003,
        month = dec,
       volume = {126},
       number = {6},
        pages = {2971-2987},
          doi = {10.1086/379553},
       adsurl = {https://ui.adsabs.harvard.edu/abs/2003AJ....126.2971V}
}

@ARTICLE{Vioque22,
       author = {{Vioque}, Miguel and {Oudmaijer}, Ren{\'e} D. and {Wichittanakom}, Chumpon and {Mendigut{\'\i}a}, Ignacio and {Baines}, Deborah and {Pani{\'c}}, Olja and {Iglesias}, Daniela and {Miley}, James and {P{\'e}rez-Mart{\'\i}nez}, Ricardo},
        title = "{Identification and Spectroscopic Characterization of 128 New Herbig Stars}",
      journal = {\apj},
         year = 2022,
        month = may,
       volume = {930},
       number = {1},
          eid = {39},
        pages = {39},
          doi = {10.3847/1538-4357/ac5c46},
archivePrefix = {arXiv},
       eprint = {2202.01234},
 primaryClass = {astro-ph.SR},
       adsurl = {https://ui.adsabs.harvard.edu/abs/2022ApJ...930...39V}
}

@ARTICLE{Wegner02,
       author = {{Wegner}, W.},
        title = "{Atlas of Interstellar Extinction Curves of OB Stars Covering the Whole Available Wavelength Range}",
      journal = {Baltic Astronomy},
         year = 2002,
        month = jan,
       volume = {11},
        pages = {1-74},
       adsurl = {https://ui.adsabs.harvard.edu/abs/2002BaltA..11....1W}
}

@ARTICLE{wise,
       author = {{Wright}, Edward L. and {Eisenhardt}, Peter R.~M. and {Mainzer}, Amy K. and {Ressler}, Michael E. and {Cutri}, Roc M. and {Jarrett}, Thomas and {Kirkpatrick}, J. Davy and {Padgett}, Deborah and {McMillan}, Robert S. and {Skrutskie}, Michael and et al.},
        title = "{The Wide-field Infrared Survey Explorer (WISE): Mission Description and Initial On-orbit Performance}",
      journal = {\aj},
         year = 2010,
        month = dec,
       volume = {140},
       number = {6},
        pages = {1868-1881},
          doi = {10.1088/0004-6256/140/6/1868},
archivePrefix = {arXiv},
       eprint = {1008.0031},
 primaryClass = {astro-ph.IM},
       adsurl = {https://ui.adsabs.harvard.edu/abs/2010AJ....140.1868W}
}

@ARTICLE{wolff04,
       author = {{Wolff}, S.~C. and {Strom}, S.~E. and {Hillenbrand}, L.~A.},
        title = "{The Angular Momentum Evolution of 0.1-10 M$_{solar}$ Stars from the Birth Line to the Main Sequence}",
      journal = {\apj},
         year = 2004,
        month = feb,
       volume = {601},
       number = {2},
        pages = {979-999},
          doi = {10.1086/380503},
archivePrefix = {arXiv},
       eprint = {astro-ph/0310280},
 primaryClass = {astro-ph},
       adsurl = {https://ui.adsabs.harvard.edu/abs/2004ApJ...601..979W}
}

@ARTICLE{Wright24,
       author = {{Wright}, Nicholas J. and {Jeffries}, R.~D. and {Jackson}, R.~J. and {Sacco}, G.~G. and {Arnold}, Becky and {Franciosini}, E. and {Gilmore}, G. and {Gonneau}, A. and {Morbidelli}, L. and {Prisinzano}, L. and et al.},
        title = "{The Gaia-ESO Survey: 3D dynamics of young groups and clusters from GES and Gaia EDR3}",
      journal = {\mnras},
         year = 2024,
        month = sep,
       volume = {533},
       number = {1},
        pages = {705-728},
          doi = {10.1093/mnras/stae1806},
archivePrefix = {arXiv},
       eprint = {2311.08358},
 primaryClass = {astro-ph.GA},
       adsurl = {https://ui.adsabs.harvard.edu/abs/2024MNRAS.533..705W}
}

@ARTICLE{Zari19,
       author = {{Zari}, E. and {Brown}, A.~G.~A. and {de Zeeuw}, P.~T.},
        title = "{Structure, kinematics, and ages of the young stellar populations in the Orion region}",
      journal = {\aap},
         year = 2019,
        month = aug,
       volume = {628},
          eid = {A123},
        pages = {A123},
          doi = {10.1051/0004-6361/201935781},
archivePrefix = {arXiv},
       eprint = {1906.07002},
 primaryClass = {astro-ph.SR},
       adsurl = {https://ui.adsabs.harvard.edu/abs/2019A&A...628A.123Z}
}

@ARTICLE{zinnecker07,
       author = {{Zinnecker}, Hans and {Yorke}, Harold W.},
        title = "{Toward Understanding Massive Star Formation}",
      journal = {\araa},
         year = 2007,
        month = sep,
       volume = {45},
       number = {1},
        pages = {481-563},
          doi = {10.1146/annurev.astro.44.051905.092549},
archivePrefix = {arXiv},
       eprint = {0707.1279},
 primaryClass = {astro-ph},
       adsurl = {https://ui.adsabs.harvard.edu/abs/2007ARA&A..45..481Z}
}

@ARTICLE{ZorecBriot91,
       author = {{Zorec}, J. and {Briot}, D.},
        title = "{Absolute magnitudes of B emission line stars : correlation between the luminosity excess and the effective temperature.}",
      journal = {\aap},
         year = 1991,
        month = may,
       volume = {245},
        pages = {150},
       adsurl = {https://ui.adsabs.harvard.edu/abs/1991A&A...245..150Z}
}





\bsp	
\label{lastpage}
\end{document}